# Engineering magnetism in hybrid organic-inorganic metal halide perovskites

*Yaiza Asensio,[a,b] Lucía Olano-Vegas,[a,b] Samuele Mattioni,[a,b] Marco Gobbi,[c,d] Fèlix Casanova,[a,d] Luis E. Hueso,*[a,d] and Beatriz Martín-García*[a,d]*

[a]*CIC nanoGUNE BRTA, 20018 Donostia-San Sebastián, Basque Country, Spain.*

[b]*Departamento de Polímeros y Materiales Avanzados: Física, Química y Tecnología, University of the Basque Country (UPV/EHU), 20018, Donostia-San Sebastian, Spain*

[c]*Materials Physics Center CSIC-UPV/EHU, 20018 Donostia-San Sebastián, Spain*

[d]*IKERBASQUE, Basque Foundation for Science, 48009 Bilbao, Spain.*

*l.hueso@nanogune.eu; b.martingarcia@nanogune.eu

## Abstract

The chemical and structural flexibility of hybrid organic-inorganic metal halide perovskites (HOIPs) provides an ideal platform for engineering not only their well-studied optical properties, but also their magnetic ones. In this review we present HOIPs from a new perspective, turning the attention to their magnetic properties and their potential as new class of on-demand low-dimensional magnetic materials. Focusing on HOIPs containing transition metals, we comprehensively present the progress that has been made in preparing, understanding and exploring magnetic HOIPs. First, we briefly introduce HOIPs in terms of composition and crystal structure and examine the synthesis protocols commonly used to prepare those showing magnetic properties. Then, we present their rich magnetic behavior and phenomenology; discuss their origin and guidelines for tuning them by changing the perovskite phase, chemical composition and dimensionality; and showcase their potential application in magneto-optoelectronics and spintronics. Finally, we describe the current challenges in the field, such as their integration into devices, as well as the emerging possibilities of moving from magnetic doping to pure transition metal-based HOIPs, which will motivate further studies in the future.

## Wider impact

This review presents our current understanding of the magnetic properties of hybrid organic-inorganic transition metal halide perovskites, as well as their modulation using crystal structure and composition as tuning knobs. Furthermore, their potential applications in a variety of fields such as magneto-optics, sensors and spin filters are described. Both these aspects are crucial to make these materials key for the future development of magnetically controlled optical and electronic devices and spin-based multifunctional technologies.

## 1. Introduction

Hybrid organic-inorganic metal halide perovskites (HOIPs) have attracted significant attention in the past years due to their remarkable optoelectronic properties, such as high photoconversion efficiency, long-range electron and hole diffusion lengths, ease of synthesis and versatility of the properties by chemical substitution, all of which have led to their successful application in photovoltaics, photodetectors and light emitting diodes (LEDs).[1–4] Although most research on HOIPs has focused on optimizing these properties, their structural and chemical versatility offers an ideal platform to engineer not only their optoelectronic behavior, but also their magnetic characteristics. Indeed, recent works[5–9] have demonstrated that by tuning elements such as the dimensionality, transition metal composition, and perovskite phase, HOIPs can exhibit a diverse range of magnetic behaviors, including ferromagnetism (FM), antiferromagnetism (AFM), and coexistence of FM/AFM. Therefore, a deeper understanding of how each of these elements contributes to the magnetic properties could open the door to the development of new functionalities and the discovery of novel magnetic phenomena.

Metal halide perovskites are compounds that follow the general formula $ABX_3$ for the three-dimensional (3D) arrangement of $[BX_6]^{4-}$ octahedra, with the peculiarity of an A-site entirely or partially occupied by small organic molecules leading to HOIPs, the focus of this review, although inorganic cations such as $Cs^+$ or $Rb^+$ can also be present. B-site is filled by metals and X-site by halogens.[10] These structures impose strict size constraints on their elements to fit within the perovskite framework. However, the dimensionality, defined as the number of spatial directions in which the octahedra formed by $[BX_6]^{4-}$ are connected, can vary from the 3D case to the two-dimensional (2D) to one-dimensional (1D) to the zero-dimensional (0D) ones.[10] As the dimensionality decreases, the A-site's size restrictions are progressively reduced, leading to a wide range of possible combinations. This broad structural and chemical flexibility of HOIPs offers substantial opportunities for fine-tuning their physical properties through simple chemical modifications and to obtain materials with several functionalities.[4,11] Among the properties that can be controlled, electrical, optical and magnetic ones stand out as the most promising, mainly due to their potential for technological advancements. In particular, the magnetic behavior of HOIPs is strongly influenced by their dimensionality, phase and composition, and even minor modifications can lead to changes in their exchange mechanism and anisotropy, giving rise to different properties.[12–14] In addition to their structural and compositional diversity, HOIPs offer another key advantage over traditional magnetic semiconductors or metals: their ease of synthesis based on solution processes that are low cost, versatile and do not require very high temperatures or pressures.[15–19]

Research on the magnetic properties of HOIPs dates back to the 1960s when de Jongh et al. studied numerous materials and established the relationship between their structure and magnetic properties.[20–24] Since then, a wide array of magnetic HOIPs[5,20–62] have been reported and, recent works have shown renewed interest in this field, highlighting their rich magnetic phenomena and their potential for exploring the interplay between magnetism and optoelectronic properties.[63] This opens exciting possibilities for developing multifunctional materials that can be employed in spintronics, magneto-optics, magnetic sensing or data storage devices, among others.[63–66]

Despite the growing interest in this area, there has been no comprehensive review dedicated to understanding and summarizing the developments related to magnetism in HOIPs. This review aims to fill this gap by providing a detailed examination of the origin of magnetic properties in these materials and how they can be tuned through dimensionality and structural modifications, the methodologies employed to synthesize magnetic HOIPs, and the strategies to exploit these materials for practical applications. We will also discuss the challenges and opportunities in this emerging field, offering insights into potential future directions and the broader impact of magnetic HOIPs in advanced technologies.

## 2. HOIP crystal structure diversity

As commented in the introduction, metal halide perovskites are characterized by a crystal structure with general formula $ABX_3$, resembling the mineral $CaTiO_3$. In the particular case of HOIPs, the A-site is occupied by organic cations, while halogens (Cl, Br or I) fill the X-site and metals occupy the B-site.[10] Typically, a single divalent metal cation is accommodated in the latter, although it can also host two different aliovalent or divalent atoms, forming the so-called double perovskites with the formula $A_2B_1B_2X_6$.[67] However, the size of the elements in these 3D structures is strictly limited by the perovskite framework. Nonetheless, this perovskite family extends beyond this dimensionality, presenting also 2D, 1D and 0D structures depending on the connectivity of the $[BX_6]^{4-}$ octahedra (corner-, edge- or face-sharing), as shown in Fig. 1 with representative examples.[68–73] As the 3D 'base' compound is progressively sliced to layers or chains, the size constrains on the organic cation, metal and halide gradually diminish, allowing for larger tunability. For instance, it is possible to accommodate large organic molecules in the A-site in 2D HOIPs, although other factors, such as the interaction between their functional groups and the inorganic framework, must also be considered for obtaining the desired material. Monoammonium ($R\text{-}NH_3^+$) and diammonium ($NH_3\text{-}R\text{-}NH_3^+$) organic molecules, where R represents the rest of the molecule, e.g. aliphatic chains or aromatic groups, are commonly used in 2D HOIPs. These cations separate the inorganic $[BX_6]^{4-}$ octahedra, which are now arranged in sheets, with the general formula $A_{n-1}B_nX_{3n+1}$, being n the number of inorganic layers and $n\rightarrow\infty$ representing the 3D counterpart.[74–78] Each

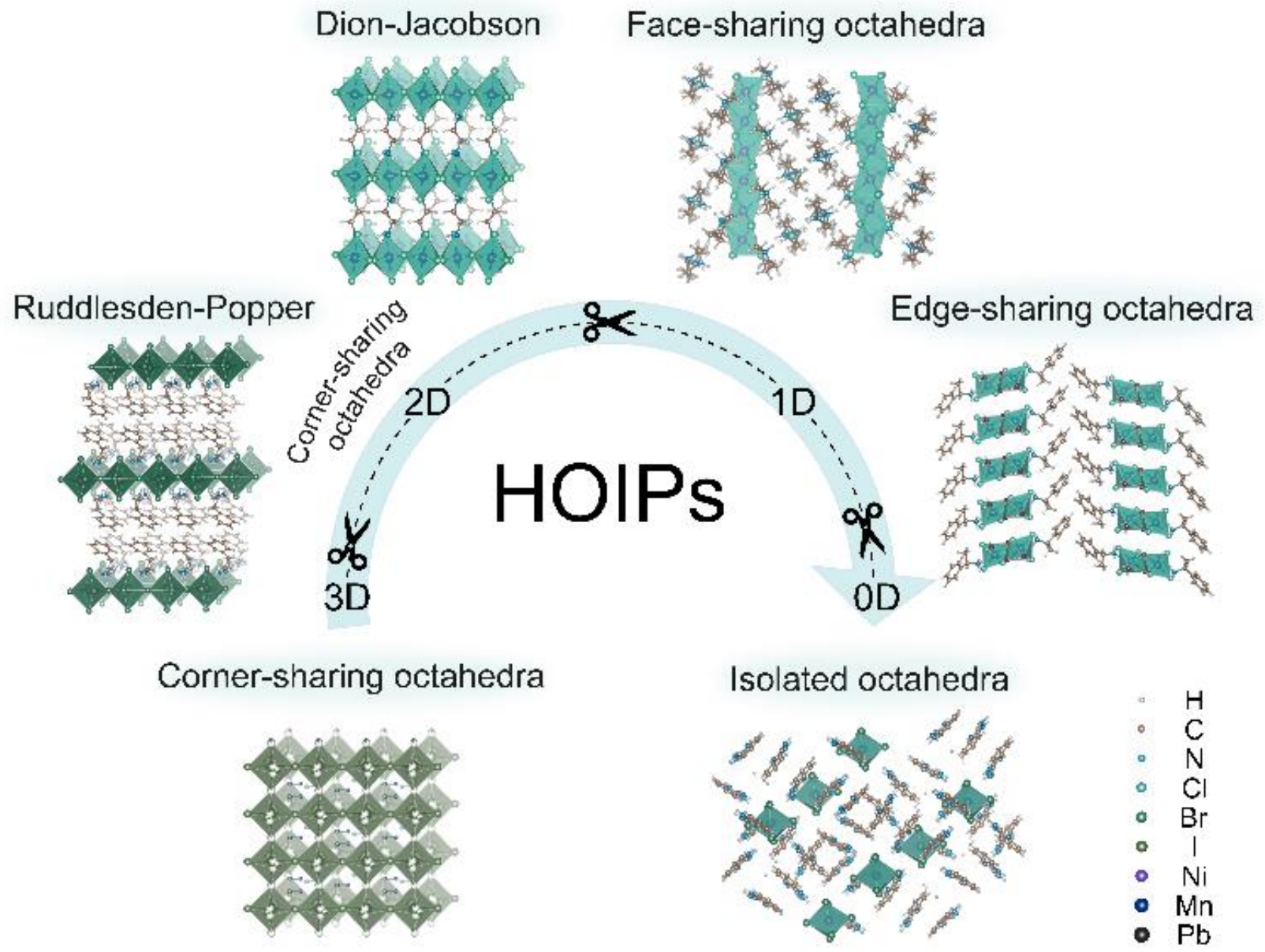

ammonium cation will lead to different hydrogen bonding schemes with the inorganic octahedra, resulting in their different conformation and orientation. In addition, these molecules give rise to two different perovskite phases: Ruddlesden-Popper (RP) and Dion-Jacobson (DJ) for mono- and diammonium molecules, respectively. 2D RP HOIPs

*Fig. 1. Crystal structures of HOIPs for their different dimensionalities. The connectivity of the inorganic octahedra ranges from corner-sharing for 3D and 2D (including both*

*RP and DJ perovskite phases) dimensionalities, edge- and face-sharing for 1D structures, to isolated octahedra in the case 0D HOIPs. 3D HOIP: $(CH_3NH_3)PbI_3$ (CCDC 1446531), 2D RP HOIP: $(C_6H_5(CH_2)_2NH_3)_2PbBr_4$ (CCDC 1903530), 2D DJ HOIP: $(NH_3(CH_2)_2NH_3)MnCl_4$ (CCDC 1149722), 1D face-sharing HOIP:$(C_{10}H_{24}C_{14}N_2Ni_xCl_x)$ (CCDC 1202166), 1D edge-sharing HOIP: $(C_8H_{11}N)MnCl_3$ (CCDC 2183841), 0D HOIP: $(C_7H_8N_3)_4MnBr_6$ (CCDC 2010937).*

with the formula $(R\text{-}NH_3)_2A_{n-1}B_nX_{3n+1}$ feature two layers of organic ammonium cations in between the inorganic ones. Each organic layer is hydrogen bonded to the halogens of the inorganic sheets on one side, and connected to the other organic layer by van del Waals (vdW) forces on the other side. In the ideal case, there is a shift between adjacent inorganic sheets, creating a staggered arrangement as shown in Fig. 1. Interestingly, the weak vdW forces allow to mechanically exfoliate these materials as happens with common 2D vdW materials.[79–82] In contrast, DJ 2D HOIPs, $(NH_3\text{-}R\text{-}NH_3)$ $A_{n-1}B_nX_{3n+1}$, have organic cations that hydrogen bond to the inorganic sheets at both ends rather than only at one end, making stronger the link between layers and therefore, more complex their exfoliation.[80] Additionally, the metal atoms are aligned directly above one another in adjacent layers, resulting in an eclipsed arrangement.

Regarding the remaining dimensionalities, restrictions progressively relax until they are not applicable altogether. In 1D HOIPs, the metal halide octahedra are connected in a linear fashion, forming a one-dimensional nanowire. These nanowires are surrounded by organic cations which can adopt either a straight or zigzag configuration, depending on whether the connection between the inorganic octahedra is face-sharing or edge-sharing, respectively. Their general chemical formula is $A_2BX_5$. Finally, in 0D HOIPs, the metal halide octahedra are entirely isolated by the surrounding organic cations, resulting in discrete, non-connected entities distributed throughout the crystal lattice without direct connection to one another. The general chemical formula for these materials is $A_4BX_6$.[83,84]

## 3. Chemical design and growth

Considering that most of the studies about magnetic HOIPs work with single bulk crystals to avoid mixture of phases, pinholes or grain boundaries affecting the properties, we focus this section on the approaches used to obtain them. Single crystal HOIPs can be synthesized exploiting different techniques, ranging from solution processes to solid-state methods or chemical/physical vapor deposition (CVD, PVD). In general, semiconductor synthesis techniques such as CVD or PVD,[85] are applicable to the fabrication of perovskite single crystals, however these methods usually require expensive and specific equipment and are more feasible for the preparation of some all-inorganic halide perovskites and thin films.[86] On the contrary, the solution-based methods show high versatility, low cost and high control on the final crystalline quality. Due to these advantages the solution approaches are the most exploited ones for the synthesis of single crystal perovskites, including magnetic HOIPs.

In general, the metal cation B source and the molecule must be dissolved into a proper solvent to get a solution from which the single crystal can grow. Commonly, metal salts, such as metal halides (e.g. $CuCl_2$, $MnCl_2$, $FeCl_2$, etc.), are used as precursors of the metal cation, while in the case of the organic cations, the molecule or the corresponding organic halide salts can be added directly. Several potential solvents can be used, either hydrohalic acids or organic solvents. In the case of the hydrohalic acids (HX = HI, HBr and HCl), they can act as solvent as well as the halide source and thus must match the type of halide in the desired perovskite. On the other hand, organic solvents, such as ethanol/methanol, N,N-dimethylformamide (DMF), dimethyl sulfoxide (DMSO), or mixtures of them, can be used. However, these solvents do not provide excess halide like HX; thus, salts with the same halide as the desired HOIPs or an external halide source must be added.[87,88]

The crystallization from solution can be thought of as a two-step process. The first step is the phase separation, or 'birth' of the new crystal, and it is called 'nucleation'. A supersaturated solution is required for crystallization to occur. Since a supersaturated solution is not at equilibrium, formation of nuclei from the solution can relieve the supersaturation and move it towards equilibrium. Once crystallization starts, however, supersaturation can be relieved by a combination of nucleation and crystal growth. In the second stage, the nucleated nuclei grow larger by the addition of solute molecules from the supersaturated solution[88,89].

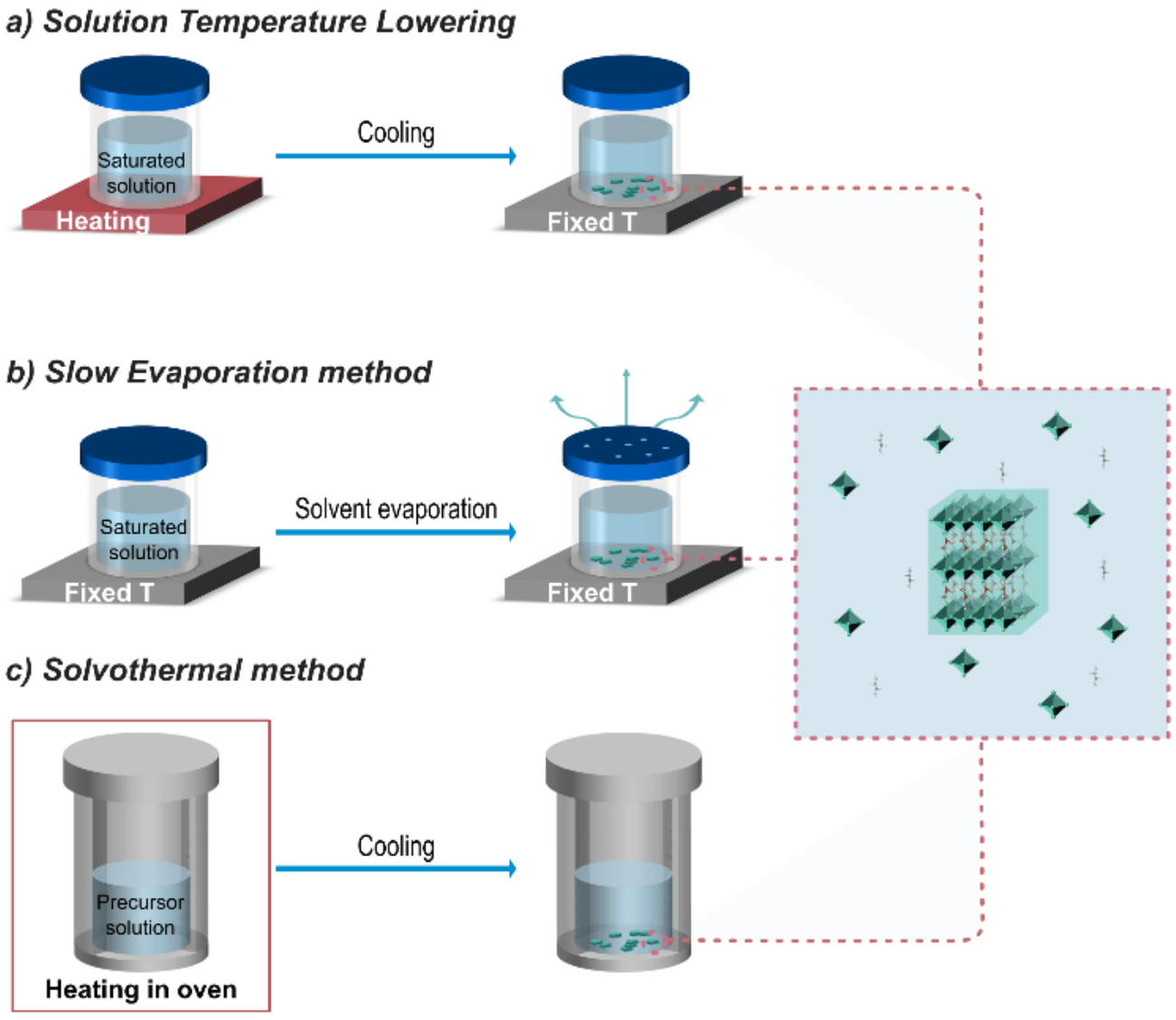

*Fig. 2. Schematics of the main solution-based process for magnetic hybrid organic-inorganic perovskite synthesis.*

In Fig. 2, sketches of the main solution-based techniques to synthesize magnetic HOIPs are shown.

### 3.1. Solution Temperature Lowering

The solution temperature lowering (STL) method exploits the fact that the solubility of the perovskite precursor decreases when lowering the solution temperature, Fig. 2a.[88] In the first stage of the

STL method, the precursors are added to the proper solvent, generally a hydrohalic acid HX but also polar solvents such as DMF, DMSO, γ-butyrolactone, acetone, ethanol or methanol, as well as their mixtures, and the system is heated to get a clear solution. In the case of hydrohalic acids, the method is versatile, since it allows synthesizing different dimensionalities as well as RP and DJ HOIP phases even reaching $n > 2$. When using polar solvents, mostly RP $n = 1$ HOIPs are obtained. Later, the solution is cooled down either at room or at a fixed temperature. The reduction of solubility induced by the cooling brings the system to supersaturation conditions, favoring the crystallization process. The cooling process can occur simply by placing the container with the saturated solution at room temperature or in a controlled way by applying a specific cooling rate. The latter allows to obtain a better control on the growth rate of the perovskite, therefore on the final crystalline quality. Therefore, the STL method is one the most exploited techniques for synthesizing magnetic HOIPs with different transition metals such as Cr, Mn, Fe or Cu as cations.[5,7,35,37–39,58,59,62,63,90–92]

### 3.2. Slow Evaporation

In the slow evaporation (SE) method, the crystals are grown from the saturated or nearly saturated solution of the compound. The solution is generally placed in a container that allows the evaporation of the solvent, e.g. a vial or beaker with a perforated cap, Fig. 2b. Through the evaporation of the solvent, the concentration of the precursor increases until the supersaturation condition is reached, and crystallization is induced.[86] Additionally, the evaporation rate could be fastened by heating. However, the use of a range of temperatures may not be desirable because the properties of the grown material may vary with temperature.[93] Together with STL, the SE method is one of the simplest approaches to obtain good quality single crystals, and similarly to STL, the SE technique is one of the most applied methods for the synthesis of magnetic HOIPs with different transition metals such as Mn, Fe or Cu as cations.[21,25,46,48–54,94–100]

### 3.3. Solvothermal method

In contrast to the other synthesis methods, the solvothermal synthetic method has the ability to increase the solubility of reactants, enhance their reactivity and prepare metastable phases which are difficult to prepare or cannot be produced by usual reactions.[101] For these reasons, the solvothermal method is widely exploited for the synthesis of magnetic double transition metal HOIPs.[6,102–104] A solution containing the precursors of the molecules and the metal cations is placed into a teflon vial and later closed in a stainless steel autoclave, Fig. 2c. The system is heated up and kept at a fixed temperature inside an oven for several hours, after which, it is then cooled down to room temperature, leading to the formation of single crystal perovskite.

Other less common approaches that one can find in literature for synthesis of magnetic HOIPs are the Antisolvent Vapor diffusion Crystallization (AVC)[105,106], where a 'poor solvent' diffuses into a 'good solvent', changing the solubility of the compound and inducing crystallization; and the Solvent Acidolysis Crystallization (SAC) generating the organic cation in the grow medium.[107] For specific systems, e.g. doped Pb perovskites, supersaturation can also be reached by an increase in the temperature (ITC - Inverse Temperature Crystallization).[108,109] Additionally, few reports can be found on solid-state synthesis of magnetic HOIPs, where the precursors are ground to form a mixture that is then heated in vacuum overnight and later cooled down at a fixed rate.[105,110]

## 4. Magnetic properties of HOIPs

### 4.1. Origin of magnetism in HOIPs

Magnetic HOIPs achieve their magnetism through the strategic incorporation of specific elements as metal cations, particularly transition metal ($Mn^{2+}$, $Fe^{2+}$, $Cr^{2+}$, $Cu^{2+}$) and rare-earth ions ($Eu^{2+}$, $Gd^{3+}$, $Sm^{3+}$) with unpaired electrons (partially filled *d* or *f* orbitals).[11] The number and arrangement of these unpaired electrons determine the magnetic order, leading to FM (aligned spins) or AFM (opposing spins) below a critical temperature. However, unlike other magnetic materials, there is no direct metal−metal contact in HOIPs. Instead, metals influence each other through the bridging halide anions and organic spacers within the HOIP structure. This indirect coupling, known as superexchange interaction, plays a crucial role in determining the strength and type of magnetism observed. According to the Goodenough-Kanamori rules,[111–113] antiferromagnetic and ferromagnetic orders can be predicted from the geometric arrangement of $[BX_6]^{4-}$ octahedra. In pure $Cu^{2+}$ or $Cr^{2+}$-based HOIPs, their octahedral frame displays a Jahn-Teller distortion due to a reduction of orbital degeneracies and crystal field effects, leading to alternatively compressed and elongated octahedral cages as shown in Fig. 3a. This configuration results in a ferromagnetic $B^{2+}$-$X^-$--$B^{2+}$ superexchange pathway through different orthogonal *d* orbitals.

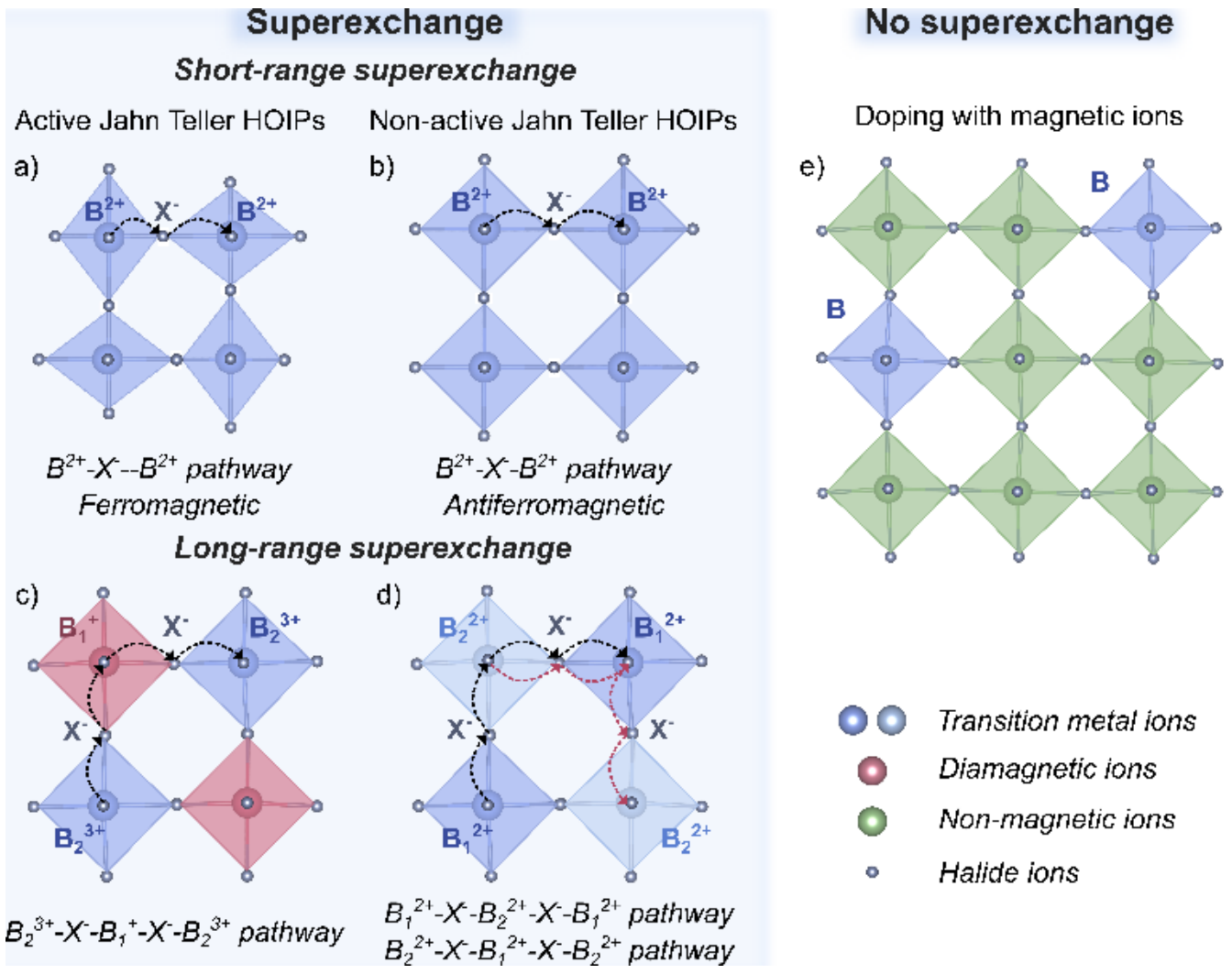

*Fig. 3. Schematic diagrams showing the presence or the absence of superexchange pathways in HOIPs. a) and b) illustrations exhibit short-range superexchange pathways for active and non-active Jahn Teller compounds, respectively. In the case of double HOIPs, long range superexchange interactions also take place as showed in c) for ($B_1^+$-$B_2^{3+}$)-based HOIPs and in d) for in ($B_1^{2+}$-$B_2^{2+}$)–based HOIPs. e) represents the absence of superexchange interactions in non-magnetic HOIPs doped with magnetic elements. Adapted with permission from ref. [7] Copyright © 2022, American Chemical Society.*

Conversely, in non-active Jahn-teller HOIPs, such as those containing $Mn^{2+}$ or $Fe^{2+}$, their linear and symmetrical B-X-B will lead to antiferromagnetic interactions following a $B^{2+}$-$X^-$-$B^{2+}$ pathway (Fig. 3b).[12]
Double HOIPs involving two metal cations ($B_1^+$-$B_2^{3+}$ or $B_1^{2+}$-$B_2^{2+}$)[6,7,13] also exhibit these superexchange interactions. However, they are now typically long-range due to an increased distance between the magnetic sites, leading to weaker magnetic exchange compared to HOIPs with a single metal cation. For double HOIPs containing a diamagnetic ion $B_1^+$ combined with a transition metal ion $B_2^{3+}$, their superexchange interactions occur via $B_2^{3+}$—$X^-$— $B_1^+$—$X^-$— $B_2^{3+}$ pathways, Fig. 3c.[13] Additionally, when two transition metals, $B_1^{2+}$-$B_2^{2+}$, are combined, these long-range superexchange interactions could proceed along either $B_1^{2+}$—$X^-$— $B_2^{2+}$—$X^-$— $B_1^{2+}$ or $B_2^{2+}$—$X^-$— $B_1^{2+}$—$X^{-}$— $B_2^{2+}$ pathways, Fig. 3d. Short-range interactions via $B_1^{2+}$-$X^{-}$-$B_2^{2+}$ pathways could also be significant, especially if there is a substantial orbital overlap between the two different metal cations, although this situation is uncommon as previously observed in double perovskites combining a transition metal from the *3d* series with another from the *4d*/*5d* one.[114,115] Furthermore, the introduction of magnetic elements as dopants into HOIPs can induce magnetism, Fig. 3e. This has been explored in Pb-based HOIPs acting as host with $Mn^{2+}$ doping. Nevertheless, the small doping concentrations usually achieved cannot induce long-range magnetic ordering in the host comparable to that observed in the previous cases.[63,64,116]

Finally, the dimensionality of HOIPs also plays a significant role regarding their magnetic properties, as its reduction implies a decrease in the number of nearest neighbors for each magnetic ion.[24] This can result in a weakening of the magnetic interactions, an enhancement of the magnetic anisotropy (which makes the material more sensitive to the direction of the applied field) or even in frustrated magnetism due to competing interactions. The combination of all these effects can lead to complex magnetic behaviors. Therefore, understanding both these factors and their interplay is crucial for tailoring magnetic properties of HOIPs.

In the following sections, we will discuss in detail how magnetism varies among different HOIPs according to their dimensionality and the type of metal cations they contain, as well as how these properties can be modulated by replacing individual components within the HOIP structure.

## 4.2. Modulation of magnetism in 2D HOIPs integrating one metal cation

A significant part of research on magnetic HOIPs has been focused on 2D materials due to their greater tunability options in comparison with their 3D counterparts. The layered structure of 2D HOIPs introduces two exchange interactions: intralayer exchange (J) and interlayer exchange (J').[24] The former represents the strong interactions governing the alignment of the magnetic moments within the inorganic layer through superexchange. On the other hand, J' represents the coupling between the magnetic moments in adjacent inorganic layers. Despite being considered a small perturbation, interlayer exchange can significantly influence the overall magnetic behavior. Several factors of the material influence the nature and strength of both J and J':

- **Perovskite phase:** each phase (RP or DJ) presents different organic spacer arrangements and octahedral alignments between layers, impacting the magnetic exchange.
- **Organic cations:** these molecules hydrogen bond with the inorganic framework in a particular way, leading to a determined orientation and conformation of the whole structure, and to a particular interlayer spacing, which in turn affects the magnetic interactions.
- **Metal:** as explained in Section 4.1, the transition metal not only introduces magnetism but also influences the rigidity of the inorganic framework, impacting the previous hydrogen bonding schemes.
- **Halogen:** this element acts as the bridge for superexchange interactions, and any change of it will affect the overall magnetism.

Therefore, understanding these factors and their interplay is crucial for controlling the magnetic properties of 2D HOIPs. Below, we explain more in detail the role of each component in determining this behavior.

**4.2.1. Role of the perovskite phase.** As previously discussed in the introduction, RP HOIPs feature inorganic layers separated by bilayers of organic cations, resulting in a 'staggered' configuration. Conversely, DJ phase presents only one layer of diammonium molecules that can hydrogen bridge both adjacent inorganic sheets, leading to an 'eclipsed' arrangement. These fundamental structural differences significantly affect the magnetism of these compounds.[11] In particular, the hydrogen bonding of monoammonium and diammonium molecules with the inorganic framework leads to different distortions of the inorganic octahedra. This modifies both metal-halide distances and the angle formed by $X^{-}$-$B^{2+}$-$X^{-}$, ultimately affecting the intralayer magnetic interactions.
Additionally, in these layered systems, the interlayer distance plays a crucial role because a larger separation weakens the exchange interactions between adjacent layers. However, the position of the metal-halide octahedra is also relevant as it can lead to different interlayer superexchange pathways. In particular, due to their in-line stacking, DJ HOIPs present a nearly linear two-halide $B^{2+}$-$X^{-}\cdots X^{-}B^{2+}$ superexchange pathway between the layers, which facilitates stronger magnetic coupling. In contrast, compounds with RP phase typically have a non-linear pathway with larger metal-metal distances for similar interlayer spacing. As a result, HOIPs with the same interlayer distances but different phases can exhibit different magnetic phenomena.

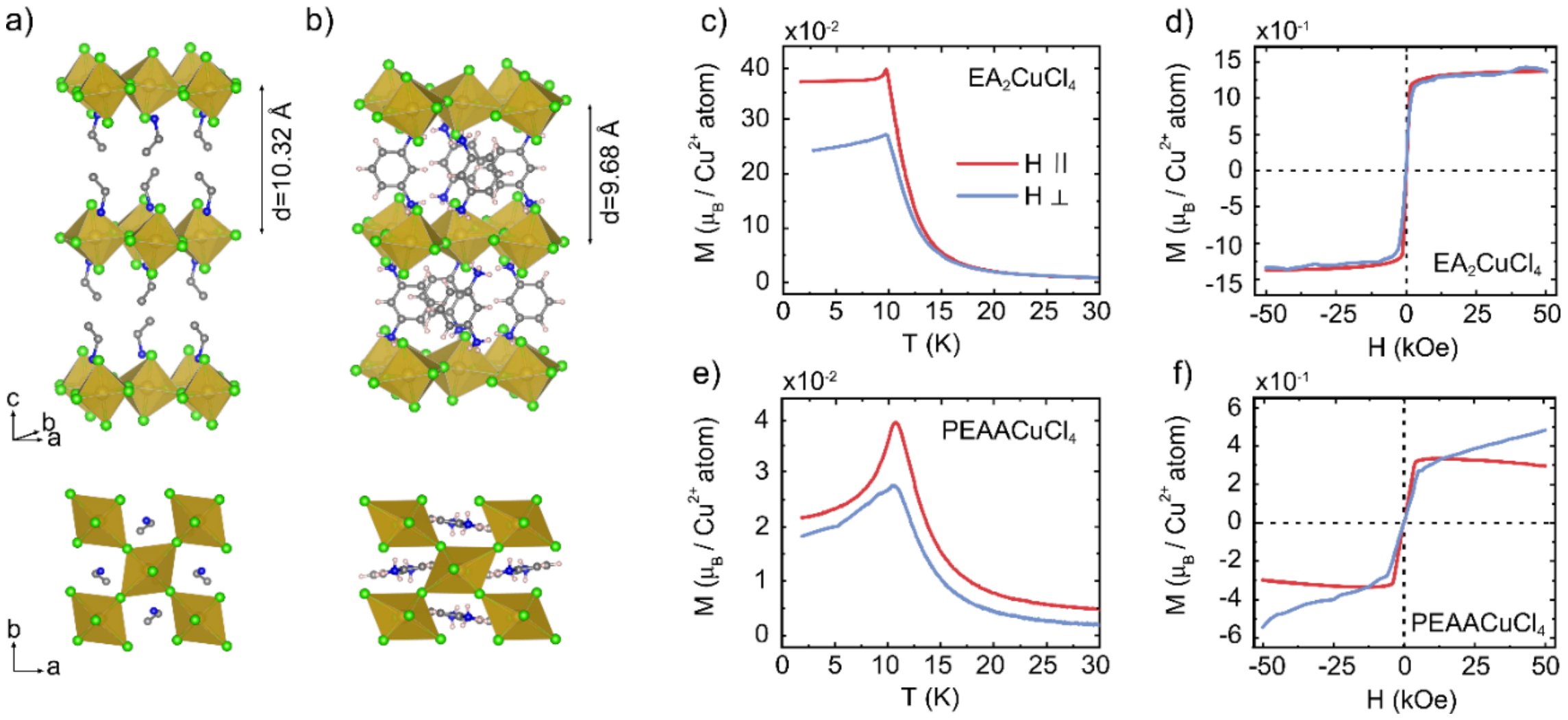

*Fig. 4. Crystal structure of a) $EA_2CuCl_4$ and b) $PEAACuCl_4$ drawn with VESTA software. Magnetization (M) versus temperature (T) at 500 Oe parallel (||, red lines) and perpendicular (⊥, blue lines) to the $[CuCl_6]^{4-}$ octahedra layers for c) $EA_2CuCl_4$ and e) $PEAACuCl_4$. Magnetization (M) versus applied field (H) at 5 K parallel (||, red lines) and perpendicular (⊥, blue lines) to the $[CuCl_6]^{4-}$ octahedra layers for d) $EA_2CuCl_4$ and f) $PEAACuCl_4$. Ref. [5] © 2022 Wiley-VCH GmbH.*

One example of this concept appears in one of our works.[5] Among the series of HOIPs studied, we analyze two compounds, $(C_2H_3NH_3)_2CuCl_4$ (ethylammonium – EA - $C_2H_3NH_3^{+}$) (RP) and $(NH_3C_6H_4NH_3)CuCl_4$ (phenylenediammonium - PEAA – $(NH_3C_6H_4NH_3)^{2+}$) (DJ), with similar interlayer distances (10.32 Å and 9.68 Å, respectively) but distinct perovskite phase (Fig. 4a-b). This difference leads to variations in the inorganic octahedra distortions, evidenced by the angles formed by the Cu-Cl-Cu atoms (~170° for $EA_2CuCl_4$ and ~160° for $PEAACuCl_4$), which translates to a higher magnetic anisotropy in the DJ phase. These structural variations have a significant impact on the magnetic behavior, that can be detected from their magnetization (M) versus temperature (T) curves for both in-plane and out-of-plane magnetic fields (Fig. 4.c-f). $EA_2CuCl_4$ exhibits a Curie temperature ($T_C$) of 10 K with the presence of a small kink for the two configurations of the field. The analysis of the exchange interaction constants reveals ferromagnetic intralayer interactions (J ~ 10 K) and much weaker antiferromagnetic interlayer interactions (J' ~ $-10^{-3}$ K), to which the emergence of the kink could be ascribed. In contrast, $PEAACuCl_4$ displays a much weaker temperature dependence of magnetization (one order of magnitude lower) with a $T_C$ around 13 K, higher than the RP-phase crystal. Additionally, the peak in the M vs T curve is much broader. While the intralayer exchange interaction remains similar to the previous compound, the interlayer interaction is significantly stronger, around $-10^{-2}$ K. This suggests that the antiferromagnetic interlayer interactions play a more dominant role in the DJ phase, influencing its overall magnetic behavior. Furthermore, from the magnetization dependance with the magnetic field (H), the DJ compound appears to achieve a saturated ferromagnetic state at higher applied fields compared to the RP phase. In summary, this example demonstrates how the structural variations between RP and DJ phases in 2D HOIPs significantly impact their magnetic behavior.

**4.2.2. Role of the organic cation.** The influence on the magnetic behavior of 2D HOIPs goes beyond just the perovskite phase (RP or DJ). The specific molecule within these categories, whether monoammonium or diammonium cation, also plays a significant role. The two key factors of the molecules that significantly affect the magnetism are the following[11,87]:

- Organic chain length. The length of the organic molecule directly impacts the spacing between the inorganic lattice

  sheets. In turn, this influences the effective dimensionality of the material with respect to its magnetic properties.
- Hydrogen bonding. The specific hydrogen bonding between the organic cations and halides influences the structural arrangement of all components within 2D HOIPs, being essential in determining the overall magnetic behavior of the material.

***4.2.2.1. Length of the organic cation.*** The distance between the inorganic layers is significantly influenced by the length of the organic cation. Longer organic chains weaken interlayer magnetic interactions in 2D HOIPs due to the larger spacing, potentially leading to lower dimensional magnetic behavior. Conversely, shorter chains strengthen these interactions, resulting even in a 3D magnetic ordering. To illustrate this correlation between the organic chain length and magnetic properties, we first focus on the extensively studied Cu-based 2D HOIPs. A wide range of compounds incorporating different organic families within the Cu-halide framework has been reported, providing a valuable dataset for investigating structure-property relationships.[20,22–34] Among them, the series of compounds $(NH_3(CH_2)_xNH_3)CuCl_4$ with x=2-10[26,27,29–31] (Fig. 5a-b) clearly shows the dimensionality reduction of the magnetic coupling while increasing the number of $CH_2$ groups (x) in the organic chain.[117]

Regarding their intralayer interactions, all the HOIPs within this series exhibit ferromagnetic behavior due to the Jahn-Teller distortion of the $[CuCl_6]^{4-}$ octahedra. The intralayer exchange constants J maintain similar values across the series, ranging from 13 to 23 K, indicating a minimal impact of the organic chain length on the intralayer magnetic coupling. In contrast, their interlayer magnetic coupling is antiferromagnetic and presents significant variations across the series. In particular, the interlayer exchange constant J' passes from values of the order of $10^{-3}$ for the highest interlayer distance (x=9), to values of the order of 10 for x=2. The ratio between the exchange constants for all the range of studied HOIPs appears in Fig. 5c. It shows how, for larger interlayer distances (higher x), the intralayer interactions are much higher than the interlayer ones, presenting 2D magnetic ordering. However, the interlayer interactions become important as the interlayer distance is reduced (x decreases), achieving the behavior of a 3D magnet for x=2.

This behavior is applicable for the same series of organics but incorporating Br as a halide[29,32,33], and also in compounds presenting a RP phase. For instance, the $(C_xH_{2x+1}NH_3)_2CuX_4$ (X=Cl, Br) series[5,20,22–24] exhibit a similar reduction of the ratio J'/J while decreasing the interlayer distance (see Table 1[20,22–34]), suggesting a general trend in 2D HOIPs. However, the J' are weaker due to the staggered arrangement of the inorganic layers. Additionally, in the case of $(C_6H_5(CH_2)_xNH_3)_2CuX_4$ series with X=Cl, Br[5,49,100,118,119] only the intralayer exchange constants are reported, showing again very low variations when changing the x. This is also revealed by Cr and Mn-based HOIPs (Table 2[35–45] and Table 3[5,21,46–55], respectively), in particular in compounds mixed with Cl and $C_xH_{2x+1}NH_3+$ [5,21,35,41,46,47,51,52] and $C_6H_5(CH_2)_xNH_3+$ [5,35,38,48–50] monoammonium molecules, confirming that organics do not have a notable influence on the intralayer interactions of 2D HOIPs.

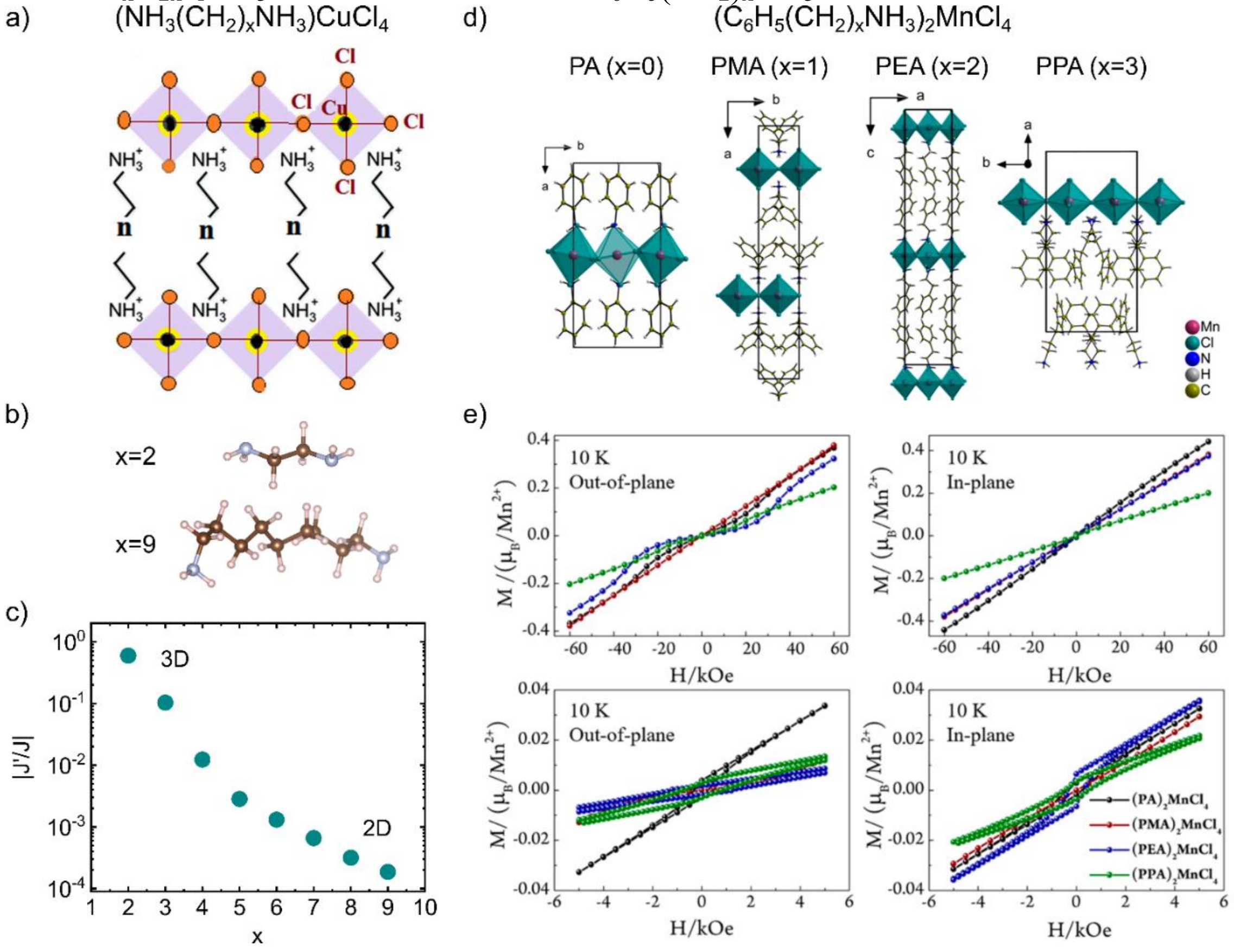

*Fig. 5. Crystal structure of a) $(NH_3(CH_2)_xNH_3)CuCl_4$ with organic molecules ranging from b) x=2 to x=9 Ref. [117] Copyright © 2017, Springer-Verlag GmbH Germany and c) absolute value of the ratio between their exchange constants |J'/J| versus x. d) Crystal structures of $PA_2MnCl_4$, $PMA_2MnCl_4$, $PEA_2MnCl_4$ and $PPA_2MnCl_4$ at 100 K and their field-dependent magnetization measured at 10 K in high and low magnetic fields along the out-of-plane and in-plane directions. Ref. [48] Copyright © 2021 The Authors.*

These results suggest that the organic chain length mainly influences the interlayer magnetic coupling in 2D HOIPs, whereas it does not have a notable impact on the intralayer interactions.

***4.2.2.2. H-bonding between the organic cations and the inorganic framework.*** Hydrogen bonding between halide anions and ammonium heads of the organic cations significantly influences the structural arrangement of 2D HOIPs. It is responsible for the alignment of neighboring inorganic sheets, the orientation of the organic molecules and the distortion of the corner-sharing $[BX_6]^{4-}$ octahedra. These structural variations, influenced by specific hydrogen bonding patterns for each organic molecule, lead to the emergence of diverse magnetic phenomena such as metamagnetism, spin-flop (SF) transitions and spin canting.

For instance, $(C_6H_5(CH_2)_xNH_3)_2MnCl_4$ series with arylamines being x=0-3 (x=0 - phenylammonium - PA, x=1 - phenylmethylammonium - PMA, x=2 PEA - phenylethylammonium, x=3 PPA - phenylpropylammonium) (Fig. 5d) reported by Septiany et al.[48], exhibit different magnetic properties depending on the organic cation.

*Table 1. Magnetic properties of Cu-based 2D HOIPs.*

| HOIP | Organic family | $n$ | Halogen | Perovskite phase | $T_C/T_N$ (K) | $J/k_B$ (K) | $J'/k_B$ (K) | $\lvert J'/J \rvert$ | Interlayer distance (Å) | Ref. |
|---|---|---|---|---|---|---|---|---|---|---|
| $(CH_3NH_3)_2CuCl_4$ | $C_nH_{2n+1}NH_3$ | 1 | Cl | RP | 8.91 | 18.20 | | 0.0006 | 9.99 | 22–24 |
| $(C_2H_5NH_3)_2CuCl_4$ | | 2 | Cl | | 10.20 | 17.20 | −0.003 | 0.0002 | 10.32 | 5, 20 and 22–24 |
| $(C_3H_7NH_3)_2CuCl_4$ | | 3 | Cl | | 7.61 | 16.00 | | 0.0001 | 12.89 | 22–24 |
| $(C_4H_9NH_3)_2CuCl_4$ | | 4 | Cl | | 7.27 | 15.40 | | 0.0001 | 15.83 | 23 and 24 |
| $(C_5H_{11}NH_3)_2CuCl_4$ | | 5 | Cl | | 7.26 | 15.90 | | 0.0001 | 17.80 | 23 and 24 |
| $(C_6H_{13}NH_3)_2CuCl_4$ | | 6 | Cl | | 7.65 | 17.10 | | 0.0001 | | 23 and 24 |
| $(C_{10}H_{21}NH_3)_2CuCl_5$ | | 7 | Cl | | 7.91 | 17.90 | | <0.00001 | 25.78 | 22 and 24 |
| $(CH_3NH_3)_2CuBr_4$ | | 1 | Br | | 15.80 | | | | 10.31 | 24 |
| $(C_2H_5NH_3)_2CuBr_4$ | | 2 | Br | | 10.85 | 19.00 | | 0.0020 | 11.44 | 23 and 24 |
| $(C_3H_7NH_3)_2CuBr_4$ | | 3 | Br | | 10.50 | 21.30 | | 0.0020 | 12.78 | 23 and 24 |
| $(C_4H_9NH_3)_2CuBr_4$ | | 4 | Br | | 11.33 | 21.90 | | 0.0010 | 14.82 | 24 |
| $(C_5H_{11}NH_3)_2CuBr_4$ | | 5 | Br | | 11.40 | 22.00 | | 0.0001 | | 24 |
| $(C_6H_5CH_2NH_3)_2CuCl_4$ | $C_6H_5(CH_2)_nNH_3$ | 1 | Cl | RP | 12.00 | 18.20 | 0.001 | <0.0001 | 18.87 | 5, 49, 100 and 119 |
| $(C_6H_5(CH_2)_2NH_3)_2CuCl_4$ | | 2 | Cl | | 9.00 | 18.80 | | | 19.69 | 100 and 119 |
| $(C_6H_5(CH_2)_3NH_3)_2CuCl_4$ | | 3 | Cl | | 7.00 | 16.70 | | | 20.44 | 100 and 119 |
| $(C_6H_5CH_2NH_3)_2CuBr_4$ | | 1 | Br | | 12.81 | 25.00 | <1.000 | <0.0400 | | 118 |
| $(C_6H_5(CH_2)_2NH_3)_2CuBr_4$ | | 2 | Br | | 12.85 | 22.70 | <0.500 | <0.0200 | | 118 |
| $(C_6H_5(CH_2)_3NH_3)_2CuBr_4$ | | 3 | Br | | 10.02 | 21.70 | <0.300 | <0.0100 | | 118 |
| $(NH_3C_2H_4COOH)_2CuCl_4$ | Others | | Cl | RP | | 13.80 | | | | 25 |
| $(NH_4)_2CuCl_4$ | | | Cl | | 11.20 | 17.00 | | 0.0032 | 8.91 | 24 |
| $((CH_3)_2CHCH_2NH_3)_2CuCl_4$ | | | Cl | | 6.50 | 14.58 | | | | 120 |
| $R/S$-$(C_6H_5CHCH_3CH_2NH_2)_2CuCl_4$ | | | Cl | | 5.00 | | | | | 121 |
| $((CH_3)_2CHCH_2NH_3)_2CuCl_4$ | | | Br | | 12.20 | 21.40 | −0.050 | | | 120 |
| $(NH_3(CH_2)_2NH_3)CuCl_4$ | $NH_3(CH_2)_nNH_3$ | 2 | Cl | DJ | 31.50 | 23.00 | −13.700 | 0.5950 | 8.11 | 26 |
| $(NH_3(CH_2)_3NH_3)CuCl_4$ | | 3 | Cl | | 14.90 | 16.50 | −1.700 | 0.1000 | 9.12 | 27 and 30 |
| $(NH_3(CH_2)_4NH_3)CuCl_4$ | | 4 | Cl | | 8.90 | 13.00 | −0.160 | 0.0100 | 9.09 | 31 |
| $(NH_3(CH_2)_5NH_3)CuCl_4$ | | 5 | Cl | | 7.60 | 14.10 | −0.040 | 0.0030 | 11.94 | 30 |
| $(NH_3(CH_2)_6NH_3)CuCl_4$ | | 6 | Cl | | 9.30 | 15.30 | −0.020 | 0.0010 | | 29 |
| $(NH_3(CH_2)_7NH_3)CuCl_4$ | | 7 | Cl | | 8.30 | 15.20 | −0.010 | 0.0007 | | 29 |
| $(NH_3(CH_2)_8NH_3)CuCl_4$ | | 8 | Cl | | 8.20 | 15.80 | −0.005 | 0.0003 | | 29 |
| $(NH_3(CH_2)_9NH_3)CuCl_4$ | | 9 | Cl | | 6.00 | 16.10 | −0.003 | 0.0002 | | 29 |
| $(NH_3(CH_2)_{10}NH_3)CuCl_4$ | | 10 | Cl | | 7.00 | 15.60 | <1.000 | <0.0200 | | 29 |
| $(NH_3(CH_2)_2NH_3)CuBr_4$ | | 2 | Br | | 72.00 | 38.20 | −68.400 | 1.8000 | | 29 and 33 |
| $(NH_3(CH_2)_3NH_3)CuBr_4$ | | 3 | Br | | 42.00 | 26.00 | −26.000 | 1.0000 | 8.60 | 32 and 33 |
| $(NH_3(CH_2)_4NH_3)CuBr_4$ | | 4 | Br | | 19.00 | 29.00 | −5.000 | 0.1700 | 8.92 | 32 and 33 |
| $(NH_3(CH_2)_5NH_3)CuBr_4$ | | 5 | Br | | 12.20 | 23.10 | −2.000 | 0.0900 | | 29 |
| $(NH_3(CH_2)_6NH_3)CuBr_4$ | | 6 | Br | | 13.00 | 20.30 | −0.100 | 0.0050 | | 29 |
| $(NH_3(CH_2)_7NH_3)CuBr_4$ | | 7 | Br | | 10.20 | 23.00 | <1.000 | <0.0500 | | 29 |
| $(NH_3(CH_2)_8NH_3)CuBr_4$ | | 8 | Br | | 12.60 | 22.10 | −0.050 | 0.0020 | | 29 |
| $(NH_3(CH_2)_9NH_3)CuBr_4$ | | 9 | Br | | 10.00 | 21.20 | <1.000 | <0.0500 | | 29 |
| $(NH_3(CH_2)_{10}NH_3)CuBr_4$ | | 10 | Br | | 10.00 | 16.80 | <1.000 | <0.0500 | | 29 |
| $(NH_3(CH_2)_2NH_3)CuCl_2Br_2$ | | 2 | Cl/Br | | 45.00 | 15.00 | −31.000 | | 8.30 | 34 |
| $((NH_3CH_2CH_2)NH_2)CuCl_4Cl$ | Others | | Cl | DJ | 11.80 | 18.70 | | 0.0028 | 11.80 | 28 |
| $(C_6N_2H_{10})CuCl_4$ | | | Cl | | 5.70 | 8.13 | −1.220 | 0.1500 | | 122 |
| $(NH_3C_6H_4NH_3)CuCl_4$ | | | Cl | | 13.00 | 12.00 | −0.025 | 0.0020 | 9.68 | 5 |
| $(NH_3C_2H_2NH_3)CuCl_4$ | | | Cl | | 36.00 | 1.30 | −1.000 | 0.7900 | 7.70 | 5 |
| $(C_6N_2H_{10})CuBr_4$ | | | Br | | 18.90 | 21.30 | −5.870 | 0.2800 | | 122 |

*Table 2. Magnetic properties of Cr-based 2D HOIPs.*

| HOIP | Organic family | $n$ | Halogen | Perovskite phase | $T_C/T_N$ (K) | $J/k_B$ (K) | Interlayer distance (Å) | Ref. |
|---|---|---|---|---|---|---|---|---|
| $(CH_3NH_3)_2CrCl_4$ | $C_nH_{2n+1}NH_3$ | 1 | Cl | RP | 42.00 | 13.00 | 9.44 | 35 and 41 |
| $(C_2H_5NH_3)_2CrCl_4$ | | 2 | Cl | | 41.00 | 10.10 | 10.71 | 35 |
| $(C_3H_7NH_3)_2CrCl_4$ | | 3 | Cl | | 39.50 | 9.30 | 12.35 | 35 |
| $(C_5H_{11}NH_3)_2CrCl_4$ | | 5 | Cl | | | 9.30 | 17.81 | 35 |
| $(CH_3NH_3)_2CrBr_4$ | | 1 | Br | | | 11.60 | | 44 |
| $(C_2H_5NH_3)_2CrBr_4$ | | 2 | Br | | | 10.00 | | 44 |
| $(C_3H_7NH_3)_2CrBr_4$ | | 3 | Br | | | 10.40 | | 44 |
| $(C_5H_{11}NH_3)_2CrBr_4$ | | 5 | Br | | | 10.70 | | 44 |
| $(C_8H_{17}NH_3)_2CrBr_4$ | | 8 | Br | | | 10.30 | | 44 |
| $(C_{12}H_{25}NH_3)_2CrBr_4$ | | 12 | Br | | | 9.80 | | 44 |
| $(C_6H_5CH_2NH_3)_2CrCl_4$ | $C_6H_5(CH_2)_nNH_3$ | 1 | Cl | RP | 37.00 | 10.60 | 15.71 | 35 and 38 |
| $(C_6H_5CH_2NH_3)_2CrBr_4$ | | 1 | Br | | 52.00 | 14.70 | 16.24 | 35, 37 and 45 |
| $(C_6H_5CH_2NH_3)_2CrBr_{3.3}Cl_{0.7}$ | | 1 | Br/Cl | | 49.00 | 12.50 | 16.10 | 35 and 36 |
| $(C_6H_5CH_2NH_3)_2CrBr_3Cl$ | | 1 | Br/Cl | | | 13.10 | | 45 |
| $(C_6H_5CH_2NH_3)_2CrBr_{1.8}Cl_{2.2}$ | | 1 | Br/Cl | | | 12.70 | | 45 |
| $(C_6H_5CH_2NH_3)_2CrBr_{0.7}Cl_{3.3}$ | | 1 | Br/Cl | | | 12.30 | | 45 |
| $(C_6H_5CH_2NH_3)_2CrBr_{3.3}I_{0.7}$ | | 1 | Br/I | | 51.00 | 11.00 | 16.15 | 35 and 39 |
| $(H_3N(CH_2)_3NH_3)_2CrCl_4$ | Others | | Cl | RP | | 10.60 | | 42 |
| $(CH_3(CH_2)_{11}NH_3)_2CrCl_4$ | | | Cl | | | 9.60 | 31.00 | 40 |
| $(NH_4)_2CrBr_4$ | | | Br | | | −5.90 | | 44 |
| $(CH_3(CH_2)_4NH_3)CrCl_4$ | Others | | Cl | DJ | | 9.30 | 17.81 | 43 |

*Table 3. Magnetic properties of Mn-based 2D HOIPs.*

| HOIP | Organic family | $n$ | Halogen | Perovskite phase | $T_C/T_N$ (K) | $J/k_B$ (K) | Interlayer distance (Å) | Spin canting | Canting degree (°) | Metamagnetism | SP transition | Ref. |
|---|---|---|---|---|---|---|---|---|---|---|---|---|
| $(CH_3NH_3)_2MnCl_4$ | $C_nH_{2n+1}NH_3$ | 1 | Cl | RP | 45.30 | −5.00 | | No | | | Yes | 21, 46 and 51 |
| $(C_2H_5NH_3)_2MnCl_4$ | | 2 | Cl | | 43.10 | −4.60 | 10.77 | Yes | 0.030 | Yes | Yes | 5, 47 and 51 |
| $(C_3H_7NH_3)_2MnCl_4$ | | 3 | Cl | | 39.20 | −4.45 | | Yes | 0.050 | No | Yes | 47 and 51 |
| $(C_5H_{11}NH_3)_2MnCl_4$ | | 5 | Cl | | 37.00 | ∼4.00 | | | | | | 52 |
| $(C_7H_{15}NH_3)_2MnCl_4$ | | 7 | Cl | | 39.00 | ∼4.00 | | | | | | 52 |
| $(C_9H_{19}NH_3)_2MnCl_4$ | | 9 | Cl | | 40.00 | ∼4.00 | | | | | | 52 |
| $(C_3H_7NH_3)_2MnBr_4$ | | 1 | Br | | 47.00 | −4.50 | | No | | No | | 47 |
| $(C_6H_5NH_3)_2MnCl_4$ | $C_6H_5(CH_2)_nNH_3$ | 0 | Cl | RP | 42.60 | −6.19 | | Yes | 0.036 | | | 48 |
| $(C_6H_5CH_2NH_3)_2MnCl_4$ | | 1 | Cl | | 44.60 | −7.47 | | No | | | | 48 |
| $(C_6H_5C_2H_4NH_3)_2MnCl_4$ | | 2 | Cl | | 45.70 | −7.65 | 19.46 | Yes | 0.060 | | Yes | 5 and 48–50 |
| $(C_6H_5C_3H_6NH_3)_2MnCl_4$ | | 3 | Cl | | 41.80 | −7.23 | | Yes | 0.029 | | | 48 |
| *R/S*-$(C_6H_5CHCH_3CH_2NH_2)_2MnCl_4$ | Others | | Br | | 3.20 | −3.64 | 17.65 | Yes | | | | 55 |
| $(NH_3(CH_2)_2NH_3)MnCl_4$ | $NH_3(CH_2)_nNH_3$ | 2 | Cl | DJ | | | 8.61 | | | | | 54 |
| $(NH_3(CH_2)_3NH_3)MnCl_4$ | | 3 | Cl | | | | 9.50 | | | | | 53 and 54 |
| $(NH_3(CH_2)_4NH_3)MnCl_4$ | | 4 | Cl | | 8.85 | 13.00 | 10.77 | | | | | 54 |
| $(NH_3(CH_2)_5NH_3)MnCl_4$ | | 5 | Cl | | | | 12.00 | | | | | 54 |
| $(NH_3C_2H_2NH_3)MnCl_4$ | Others | | Cl | DJ | 80.00 | −10.20 | 8.30 | | | | Yes | 5 |

*Table 4. Magnetic properties of Fe-based 2D HOIPs.*

| HOIP | Organic family | $n$ | Halogen | Perovskite phase | $T_C/T_N$ (K) | $J/k_B$ (K) | Interlayer distance (Å) | Spin canting | Canting degree (°) | Metamagnetism | SP transition | Ref. |
|---|---|---|---|---|---|---|---|---|---|---|---|---|
| $(CH_3NH_3)_2FeCl_4$ | $C_nH_{2n+1}NH_3$ | 1 | Cl | RP | 95.00 | | 9.50 | Hidden | 1.4 | Yes | No | 56, 57, 59 and 61 |
| $(C_2H_5NH_3)_2FeCl_4$ | | 2 | Cl | | 110.00 | | 10.50 | Yes | 0.63 | No | No | 57 and 61 |
| $(C_3H_7NH_3)_2FeCl_4$ | | 3 | Cl | | 90.00 | | | | | | | 58 and 61 |
| $(C_4H_9NH_3)_2FeCl_4$ | | 4 | Cl | | 90.00 | | | Yes | | | | 58 |
| $(C_6H_5CH_2NH_3)_2FeCl_4$ | $C_6H_5(CH_2)_nNH_3$ | 1 | Cl | RP | 73.00 | 78 | | | | | | 56 and 61 |
| $(C_6H_5C_2H_4NH_3)_2FeCl_4$ | | 2 | Cl | | 98.00 | | 19.50 | Yes | 0.53 | | | 62 |
| $(CH_3NH_3)_2FeBr_2Cl_2$ | Others | | Cl | RP | 100.00 | | | Yes | | No | Yes | 60 |
| $(NH_3(CH_2)_2NH_3)FeCl_4$ | $NH_3(CH_2)_nNH_3$ | 2 | Cl | DJ | | | | | | | | 53 |

Their M vs out-of-plane H measurements (Fig. 5e) reveal a SP transition only for the PEA-based HOIP, characterized by a change in slope around 25 kOe. Additionally, hysteresis loops observed at low fields in both this characterization and in M versus in-plane H measurements suggest spin canting in compounds with PA, PEA and PPA organic cations. Spin canting is a magnetic phenomenon in which spins are tilted a few degrees angle from their axis, known as canting angle. This phenomenon arises from the antisymmetric Dzyaloshinsky– Moriya (DM) interactions, caused by the tilting of the $[MnCl_6]^{4-}$ octahedra due to H-bonding between the inorganic layer and the organic part. Spin canting prevents the perfect antiparallel alignment of the spins on neighboring metal ions within an AFM layer, generating residual spins along the canted direction and inducing weak FM.[120] Their calculated canting angles are 0.036, 0.060 and 0.029 degrees, respectively, showing no correlation between the length of the organic cation.
The series of HOIPs $(C_xH_{2x+1}NH_3)_2MnCl_4$ with aliphatic chains being x=1-3[5,21,46,47,51] also demonstrates the influence of the hydrogen bonding on the magnetic behavior. In this case, all the compounds exhibit SF transitions. This phenomenon appears in AFM systems with weak anisotropy when a magnetic field is applied parallel to the preferred axis of sublattice magnetization. Under these conditions, a competition arises between the strength of the external field and the internal exchange field. In such systems, when the field reaches a critical value, the antiparallel magnetizations (spins) of the two sublattices change their orientation from the easy axis to a direction perpendicular to it, while maintaining their antiparallel alignment. This transition is referred to as SF.[120] In addition, x=1 and x=2 perovskites have spin canting, with degrees in the same range as the previous ones: 0.030 and 0.050 degrees, respectively. Notably, x=2 also presents a metamagnetic behavior, where one of the spin sublattices rotates to align parallel to the other when a strong enough magnetic field is applied along the preferred axis, overcoming the internal exchange interactions and inducing FM.[120]
However, these emergent magnetic properties can vary when the metal cation is modified. For example, when $Fe^{2+}$ replaces $Mn^{2+}$ in the previous HOIP family with aliphatic amines, $C_xH_{2x+1}NH_3^+$ [56,57,59,61], spin canting is again reported for x=1, 2 but with much larger canting angles: 1.4 and 0.63 degrees, respectively (Table 4[56–62]). In contrast, neither of them exhibits now a SF transition and the metamagnetism is now present in x=1 instead of in x=2 as in the Mn-based compound. Another Fe-based HOIP that maintains its spin canting is the one containing PEA molecules[62], which is again stronger than its Mn counterpart (canting angle of 0.53 degrees). Interestingly, this perovskite's ferroelastic nature can change its structure in response to temperature or mechanical stress, what in turn can affect the magnetic properties. In particular, uniaxial stress in this material can switch the spin canting direction. The pristine crystal exhibits a magnetic hysteresis loop along the *a*-axis, but this vanishes when applying uniaxial mechanical stress along the *b*-axis due to axis swapping, revealing a coupling between the antisymmetric Dzyaloshinskii–Moriya interaction and ferroelasticity in $PEA_2FeCl_4$.

Regarding 2D HOIPs presenting a DJ phase, limited research has been conducted to gain insight into hydrogen bonding in diammonium molecules. Only $(NH_3(CH_2)_2NH_3)MnCl_4$ (ethylenediammonium – EDA - $(NH_3(CH_2)_2NH_3)^{2+}$) is reported to show a SF transition less pronounced compared to those observed in RP HOIPs.[5]

All these findings suggest that the unique hydrogen bonding patterns for each molecule can impact the magnetic behavior. Nevertheless, a direct correlation between specific magnetic phenomena and particular features of the organic molecules is not evident.

**4.2.3. Role of the halogen.** The type of halogen significantly influences the magnetic parameters, such as the intralayer and interlayer exchange constants, Curie-Weiss temperature ($\theta_{CW}$) and $T_C$. As mentioned earlier, the organic cation mainly determines the separation between the inorganic layers and the magnetic exchange interactions occurs via halogen-mediated pathways. Given a fixed interlayer distance, larger size halogens result in shorter halogen-halogen distances, enhancing their interactions and increasing the orbital overlap between metal ions. This typically leads to larger intralayer and interlayer exchange constants.[121,122] In consequence, if for example Cl is partially or completely replaced by Br, which has a larger atomic radius, the magnetic exchange interactions are generally strengthened.

Most of the previously discussed Cu-based HOIPs incorporating different organic series were synthesized using Cl and Br as halogens. Among the perovskites with the RP phase, it can be observed in Table 1 how compounds with $C_xH_{2x+1}NH_3^+$ [5,20,22–24] (aliphatic) and $C_6H_5(CH_2)_xNH_3^+$ [5,49,100,118,119] (aryl) ammonium cations combined with Br exhibit slightly larger $T_C$ and J, as well as a lower ratio between interlayer and intralayer interactions, indicating higher J' values compared to the Cl-based counterparts. However, more notable differences emerge in 2D HOIPs with the DJ phase, especially within the $(NH_3(CH_2)_xNH_3)^{2+}$ series.[26,27,29–33] When this 2D HOIP is x=2 and contains Cl, its intralayer and interlayer exchange constants are 23.0 K and -13.7 K, respectively. Conversely, when the halogen used is Br, these two values increase to 38.2 K and -68.4 K, resulting in unusual stronger interlayer interactions compared to the intralayer ones. Additionally, $T_C$ also rises from 31.5 K to 72.0 K. Also, a compound mixing both halogens is reported[34], $(NH_3(CH_2)_2NH_3)CuCl_2Br_2$, presenting intermediate $T_C$ and J' values (45 K and -31 K) but, surprisingly, having a value of 15 K for the interlayer interactions.

Nevertheless, when the metal is Cr, some discrepancies arise. For instance, compounds with the $C_xH_{2x+1}NH_3^+$ cation series with x=1[35,41] mixed with Cl halogen exhibit a slightly higher value for J (13.0 K) than the Br analogue (11.6 K).[44] Moreover, for this metal combined with $C_6H_5(CH_2)_xNH_3^+$ (x=1), six compounds varying the Cl:Br ratio have been reported.[35–38,45] Generally, as expected, increasing the amount of Br (and reducing the amount of Cl) leads to a rise in the intralayer exchange constant, ranging from 10.6 K for pure Cl-based to 14.7 K for pure Br-based HOIP (Table 3). However, $(C_6H_5CH_2NH_3)_2CrBr_{3.3}Cl_{0.7}$ deviates from this trend, presenting a lower J value (12.5 K) compared to $(C_6H_5CH_2NH_3)_2CrBr_3Cl$ (13.1 K). Regarding their $T_C$, only values for the pure 2D HOIPs and $(C_6H_5CH_2NH_3)_2CrBr_{3.3}Cl_{0.7}$ are reported, following the same trend as Cu-based HOIPs. The compound containing only Br exhibits the highest $T_C$ (52 K), followed by the Br/Cl mixture (49 K) and finally, pure Cl-based HOIP presents the lowest one (37 K). Additionally, another work reported this 2D HOIP with the presence of Br and I[35,39], $(C_6H_5CH_2NH_3)_2CrBr_{3.3}I_{0.7}$. Due to the larger atomic radius of I, one could expect larger $T_C$ and J values for this compound than for $(C_6H_5CH_2NH_3)_2CrBr_4$, however, it presents a much lower J value, 11.0 K and a similar $T_C$, 51.0 K, highlighting the complex interplay of factors influencing magnetic properties in these materials.

### 4.3. Modulation of magnetism in HOIPs of other dimensionalities integrating one metal cation

The dimensionality of the HOIPs system can affect the final magnetic properties of the material, since a change in the connectivity of the magnetic ions can influence significantly the magnetic coupling inside the material. Similarly to the 2D counterparts, the magnetic behaviour of 0D, 1D and 3D HOIPs can as well be tuned for example by the choice of transition metal cation and the organic ligand.

Regarding 3D magnetic HOIPs, Pb-free $(CH_3NH_3)FeCl_3$ show a typical paramagnetic behavior, different from the expected behavior of 2D $Fe^{2+}$ based HOIPs[123]. Numerous reports of 3D HOIPs can be found in literature related to Mn or Fe doped Pb halide perovskites, which will be detailed in the next section.

Moving toward 1D HOIPs, Lee et al.[105,106,110] reported the magnetic properties of $ABCl_3$ systems, studying the different effects of changing the organic cation A in quasi 1D perovskite chains system containing $Ni^{2+}$, $Fe^{2+}$ or $Co^{2+}$, with face-sharing octahedra. $ANiCl_3$ compounds[106] (A = $N(CH_3)_4^+$, $CH_3NH_3^+$, $(CH_3)_2NH_2^+$, $C(NH_2)_3^+$, and $CH(NH_2)_2^+$) follow the superexchange model, and the magnetic correlations within individual Ni−Cl chains turn from AFM to FM at an average Ni−Cl−Ni angle of approximately 78°. In contrast, the 1D chains of $ACoCl_3$ compounds[105] (A = $CH_3NH_3^+$, $CH(NH_2)_2^+$, $C(NH_2)_3^+$), independently on the choice of the molecule, couple antiferromagnetically, but as evidenced by the lack of a relationship between the Co−Cl−Co angle and the corresponding intrachain coupling constant, the strength of the intrachain coupling cannot be forecasted by the superexchange model. However, there is a significant dependence on the interchain Co−Co distance, which directly affects the temperature at which long-range magnetic order is exhibited. While the $AFeCl_3$ perovskites[110] (A = $CH_3NH_3^+$, $CH(NH_2)_2^+$, $C(NH_2)_3^+$, $C_3H_5N_2^+$) display intrachain ferromagnetic interactions without a direct trend between the Fe−Cl−Fe angle and the resulting coupling strength. the $AFeCl_3$ perovskites[110] (A = $CH_3NH_3^+$, $CH(NH_2)_2^+$, $C(NH_2)_3^+$, $C_3H_5N_2^+$) display intrachain ferromagnetic interactions without a direct trend between the Fe−Cl−Fe angle and the resulting coupling strength.

1D perovskite of trimethylchloromethylammonium (TMCM – $ClCH_2NH_3^+$) chromium chloride (TMCM–$CrCl_3$), with infinite linear chains of face-sharing $CrCl_6$ octahedra, separated by TMCM cations have been studied by Ai et al.[124] The material exhibits a spin-canted antiferromagnetic behaviour with strong antiferromagnetic coupling and $T_N$ at 4.8 K. At the same time, 1D $Cu^{2+}$ based HOIPs have been studied as well. The magnetic properties of $(C_5H_8N_3)CuCl_3$ (being $C_5H_8N_3$ = 2-Amino-4-methylpyrimidinium), can be described by the model of S = ½ antiferromagnetic dimers with $J/k_B$ = - 122.7 K[125], where the exchange interactions can be evaluated using the results of Hatfield and Hodgson et al.[126] While, on the opposite, 1D hybrids such as $(C_5H_{14}N_2)CuCl_4$ (hexahydro-1,4-diazepinium as organic cation) exhibit paramagnetic behavior due to the cancellation of ferro and anti-ferromagnetic components of superexchange interactions between the magnetic centers in the compound.[127]

Looking at 1D Mn-based HOIPs, Taniguchi et al. explored [(S)/(R)-MPEA]$_2$[$MnCl_4(H_2O)$][55] with chiral R-/S-β-methylphenethylammonium (MPEA) as organic cations, showing that these compounds present antiferromagnetic interactions through the bridging single Cl ion, along individual chains of the $Mn^{2+}$ ions. Additionally, spontaneous weak FM below 3.2 K was observed and ascribed to the canting of the antiferromagnetic spins induced by the Dzyaloshinskii–Moriya interactions.

Moving from 1D to 0D HOIPs, the magnetic properties of $(H_2DABCO)MX_4{\cdot}cH_2O$ (DABCO = 1,4-diazabicyclo[2.2.2]octane), M = Mn and Cu; X = Cl and Br; *c* = 0, 1, and 4) has been reported by Panda et al.[128] All the compounds, showing a structure with isolated octahedra, are paramagnetic in nature, except for the Cu-based halides that exhibit dominant ferromagnetic interactions. Additional low-dimensional Ru based compounds studied by Vishnoi et al.[102], e.g. $(CH_3NH_3)_2RuCl_6$, display 0D structure with $[RuCl_6]^{2-}$ octahedra, showing a typical single-ion behavior without the influence of exchange interactions.

Finally, Zheng et al.[129] have shown the effect of the dimensionality on the magnetic properties of divalent Mn-based metal halides. Compared to the canted antiferromagnetic behavior with Neel temperature ($T_N$) at 45 K of the 2D compounds, the 1D system displays antiferromagnetic behavior with much lower $T_N$ at 5 K and absence of hysteresis in the M-H loop. Oppositely, the 0D perovskite presents the typical paramagnetic nature due to the isolated 0D $[MnCl_4]^{2-}$ tetrahedral framework. Additionally, the halide Mn hybrids show nanosecond-scale spin coherence times that satisfy the relation 2D > 1D > 0D at high-temperature regime.

These results confirm how the different dimensionality of HOIP can play a crucial role in determining the final magnetic properties of the material, and how within each dimensionality, the magnetism is affected by the chemical design.

## 4.4. Modulation of magnetism in HOIPs integrating two metal cations

**4.4.1. Double HOIPs.** HOIPs with one metal cation already represent a rich chemical and structural diversity framework that allow the opportunity to incorporate different magnetic spins. However, double perovskites with the general formula $A_2B_1B_2X_6$ for the 3D or $A_4B_1B_2X_8$ for the 2D counterpart can provide additional composition diversity that in turn may result in promising new and tunable properties.[130] As mentioned in Section 4.1, both $B_1^{+}$-$B_2^{3+}$ and $B_1^{2+}$-$B_2^{2+}$ combinations are possible.

Regarding $B_1^{+}$-$B_2^{3+}$ based double HOIPs, several works report changes in their magnetic behaviour depending on the choice of the diamagnetic $B_1^{+}$, which affects the long-range superexchange interaction $B_2^{3+}$—$X^{-}$— $B_1^{+}$—$X^{-}$— $B2^{3+}$. This has been demonstrated in the work by Xue et al.[7], where they investigated the magnetism of 2D Fe-Cl based double perovskites [$A_4B_1B_2Cl_8$, $B_1^{+}$ = $Ag^{+}$/$Na^{+}$; $B_2^{3+}$= $Fe^{3+}$/$In^{3+}$], reported in Fig. 6a. They demonstrate that the nearest $Fe^{3+}$–$Fe^{3+}$ centers couple antiferromagnetically and the magnetic coupling strength can readily be tuned by the bridging diamagnetic $B^{+}$. The studied $(C_6H_5(CH_2)_2NH_3)_4AgFeCl_8$ compared to the $(C_6H_5(CH_2)_2NH_3)_4NaFeCl_8$, presents higher magnetic coupling as a result from the differences in orbital hybridization, where $Ag^{+}$ *4d* orbitals strongly hybridize with Cl *3p* orbitals, promoting the long-range superexchange interactions. Additionally, when organic cations are altered in the series of Ag-Fe-Cl, $\theta_{CW}$ become less negative from ($C_6H_5(CH_2)_2NH_3^{+}$, PEA) (−10.7 K) to $(NH_3(CH_2)_4NH_3)^{2+}$ (1,4 – butanediammonium , BDA) (−3.31 K) and to chiral R-MPEA (−1.37 K), suggesting a reduced antiferromagnetic coupling, as a result of the structural distortion in the inorganic frameworks due to the particular H-bond between each molecule and the inorganic framework ($d_{Fe-Fe}$: PEA < BDA < R-MPEA).

Magnetic systems with different transition metal $B_1^{3+}$ can be found in literature. Binwal et al.[104] reported new chloride double perovskites with $Na^+/Ag^+$ and $Mo^{3+}$ metal ions and different dimensionality. Similarly to the Fe-based perovskites, the magnetic coupling of these systems (antiferromagnetic) primarily takes place through the $–Mo^{3+}–Cl–M^+–Cl–Mo^{3+}–$ pathway, the strength of which depends not only on the nearest-neighbor Mo/Mo distance, but also on the geometry of the magnetic sub-lattice. The 1D compound $(CH_3NH_3)_2AgMoCl_6$ features an antiferromagnetic transition temperature $T_N$ of 2.35 K, while the calculated frustration index of1.2 indicates that this 1D system is magnetically non-frustrating. Compared to the 2D system (1,4-($NH_3(CH_2)_4NH_3$))$_2$AgMo$Cl_8$, it displays higher antiferromagnetic exchange due to reduced distance between neighbor-spins. Similarly, the compounds containing $Na^+$-$Mo^{3+}$ are ordered antiferromagnetically below $T_N$ (5.2-6.8 K). In the $(CH_3NH_3)_2NaMoCl_{(6-x)}Br_x$ series of compounds[131], the strength of exchange interaction increases with increasing the Br content $x$.

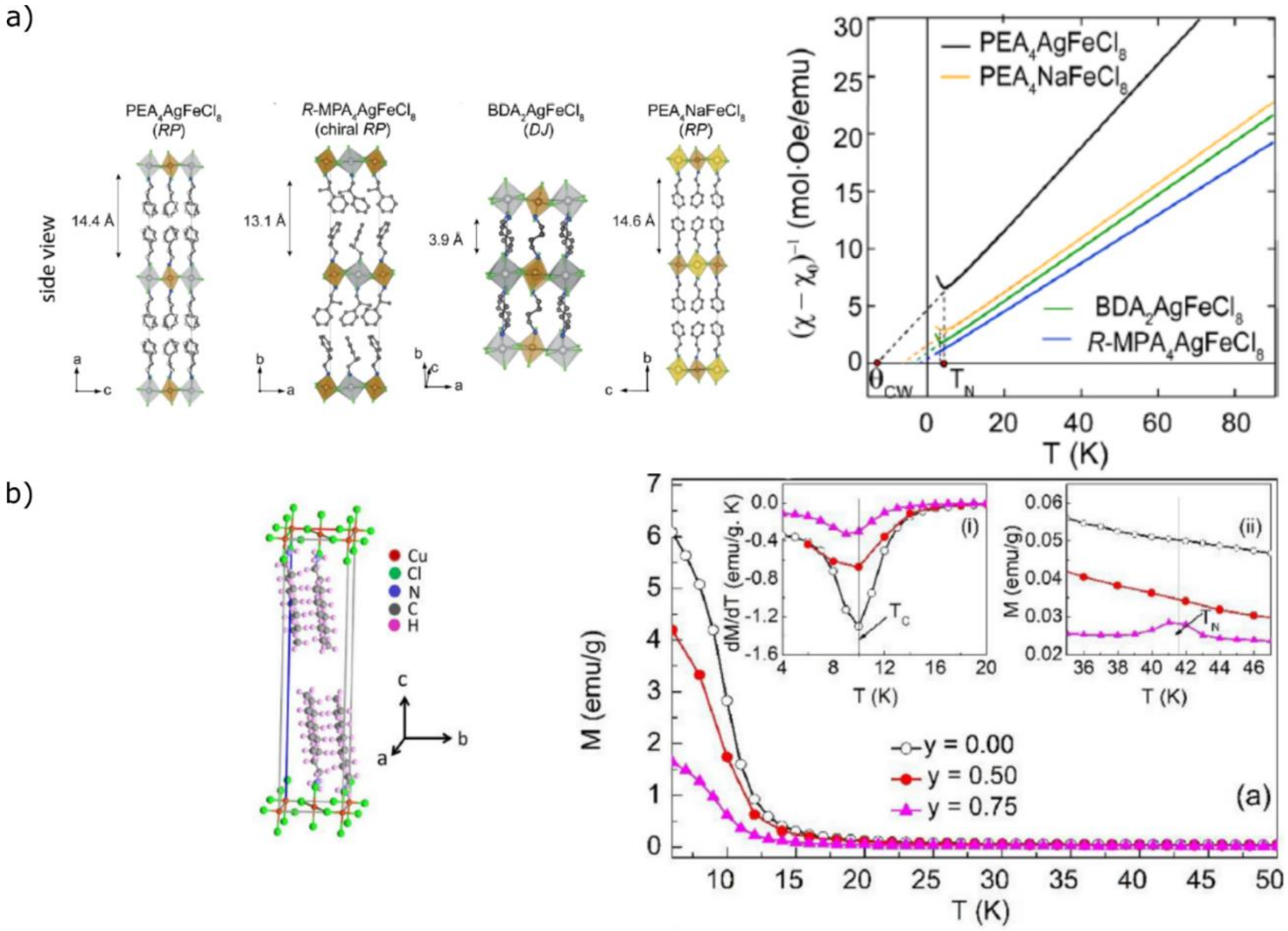

*Fig. 6. a) Temperature dependence of the inverse magnetic susceptibility of $PEA_4AgFeCl_8$, $PEA_4NaFeCl_8$, $BDA_2AgFeCl_8$, and $(R\text{-}MPA)_4AgFeCl_8$ Reprinted with permission from ref. [7] Copyright © 2022 American Chemical Society. b) M-T measurement data at fixed applied field of 400 Oe for $(C_{12}H_{25}NH_3)_2Cu_1$ y $Mn_yCl_4$ (y = 0, 0.5 and 0.75). Ref. [94] © 2019 Elsevier Inc. All rights reserved.*

The weak antiferromagnetic coupling between $Mo^{3+}$ ions is further confirmed by Vishnoi et al.[6] The group additionally studied $Ru^{3+}$ based double perovskites. Differently, the magnetic behavior of these compounds is more complex due to the unquenched orbital angular momentum of the low spin $t_{2g}^5$ of $Ru^{3+}$. The 1D $(CH_3NH_3)_2NaRuCl_6$ does not follow the behavior described by Kotani in which compounds showing isolated and undistorted $[RuCl_6]^{3-}$ octahedra exhibit a single ion behavior without the influence of exchange interaction. The lower moment of the double HOIPs in comparison with the ideal Kotani prediction might arise from increased exchange interactions resulting from the higher connectivity between $RuCl_6$ within the chains.[102]

Few studies are present for double perovskite combining metal cations with the same oxidation state $B_1^{2+}$-$B_2^{2+}$. 2D layered $(CH_3(CH_2)_{11}NH_3)_2Cu_{1-y}Mn_yCl_4$ hybrid systems at different compositions (y = 0.0, 0.5, 0.75 and 1.0) have been reported by Bochalya et al.[94] The magnetic M-T curves are shown in Fig. 6b. In this study, while crystals with y = 0, 0.5 and 0.75 show typical paramagnetic to ferromagnetic transition at $T_c$ ~10 K, compounds with y = 1, present the usual antiferromagnetic interactions. The system with composition y = 0.75, i.e., $(CH_3(CH_2)_{11}NH_3)_2Cu_{0.25}Mn_{0.75}Cl_4$, shows two transitions at temperatures of ~ 9 K and ~ 41.5 K. Although this suggests the coexistence of ferromagnetic and antiferromagnetic orders in y = 0.75, what happens is that both the $(CH_3(CH_2)_{11}NH_3)_2CuCl_4$ and $(CH_3(CH_2)_{11}NH_3)_2MnCl_4$ crystalline phases are present, and thus, their magnetic behavior.

More complex systems containing three metal cations with different oxidation states have also been reported. Connor et al.[132] synthesized a double layered perovskite $Cu^+$-$In^{3+}$ with half of the metal sites replaced by $Cu^{2+}$. The alloyed perovskite shows weak ferromagnetic interactions between local $Cu^{2+}$ moments with no long-range order down to T = 2 K, without the formation of any $Cu^{2+}$ cluster.

**4.4.1. Doped HOIPs.** The magnetism in non-magnetic HOIPs can be induced by doping of transition metal ions or rare earth ions into the perovskite lattice. Among these kinds of materials, Pb-based HOIPs are the most studied. Numerous reports on $Mn^{2+}$ doping can be found in literature both for 3D[63,116,133] and 2D[64] HOIPs.

Rajamanickam et al.[116] reported how the magnetic properties changes at different Mn amount for the $(CH_3NH_3)Pb_{1-y}Mn_yI_3$ films at room temperature. The magnetization as function of magnetic field (Fig. 7a) shows that the pristine (y = 0) and Mn doped films (y = 0.01) exhibit negligible M; nevertheless, M emerges for films with y > 0.03, showing ferromagnetic behavior. A drastic enhancement of the saturation magnetization was found for y = 0.15 (Fig. 7b). The likely presence of $Mn^{2+}$−$I^-$−$Mn^{3+}$ motifs in the sample suggests a plausible explanation for the ferromagnetic coupling observed between Mn ions. That is, the double exchange mechanism proposed by Zener[133] based on FM arising from the indirect interaction between ions with multiple oxidation states, and mediated by the halogen. However, these $(CH_3NH_3)Pb_{1-y}Mn_yI_3$ films should be considered as a particular case. The effect of $Fe^{2+}$ on the magnetism of Pb-based HOIPs has been reported as well in literature both from theoretical[134] and experimental point of view.[98,108,109]

Fu et al. [109] showed that depending on the composition, different magnetic behavior can be expected. $(CH_3NH_3)(Pb_{0.8}Fe_{0.2})I_3$ is paramagnetic, while $(CH_3NH_3)(Pb_{0.4}Fe_{0.6})I_3$ is a low-temperature antiferromagnetic with a $T_N$ of 23 K. Paramagnetic behavior has been reported as well by Bonadio et al.[98] for Fe-doped microwires halide perovskites. On the contrary, Pb-based HOIPS, e,g $(CH_3NH_3)PbCl_3$, co-doped with magnetic ($Fe^{2+}$) and aliovalent ($Bi^{3+}$) metal ions, can present ferromagnetic interactions below 12 K when the concentration of $Fe^{2+}$ and $Bi^{3+}$ reaches respectively 3% and 2.9%[108].

Therefore, these results show how it is possible to take advantage of the introduction of different metal cations in the HOIPs for tuning their magnetic properties.

## 5. Applications

The integration of transition metals into metal halide perovskites is unlocking new magnetic properties that are key to developing advanced applications in magnetism. As explained in the previous sections, by introducing transition metals into HOIPs, researchers have been able to induce magnetic properties and phenomena such as FM, spin filtering, and magneto-optoelectronic effects, which are essential for devices that leverage magnetic functionality at the proof-of-concept or lab scale. Magnetic applications of HOIPs containing transition metal cations are still in the early stages and most studies focus on doping well-established Pb-based HOIPs, like $(CH_3NH_3)PbI_3$ and $(C_6H_5(CH_2)_2NH_3)_2PbX_4$ with the purpose of comparing their behavior before and after the inclusion of magnetic counterparts. Nonetheless, there are also few reports covering pure transition-metal-based HOIPs, such as $(MBA)_2CuCl_4$ (MBA = methylbenzylammonium) and $(CH_3NH_3)_2MnCl_4$.

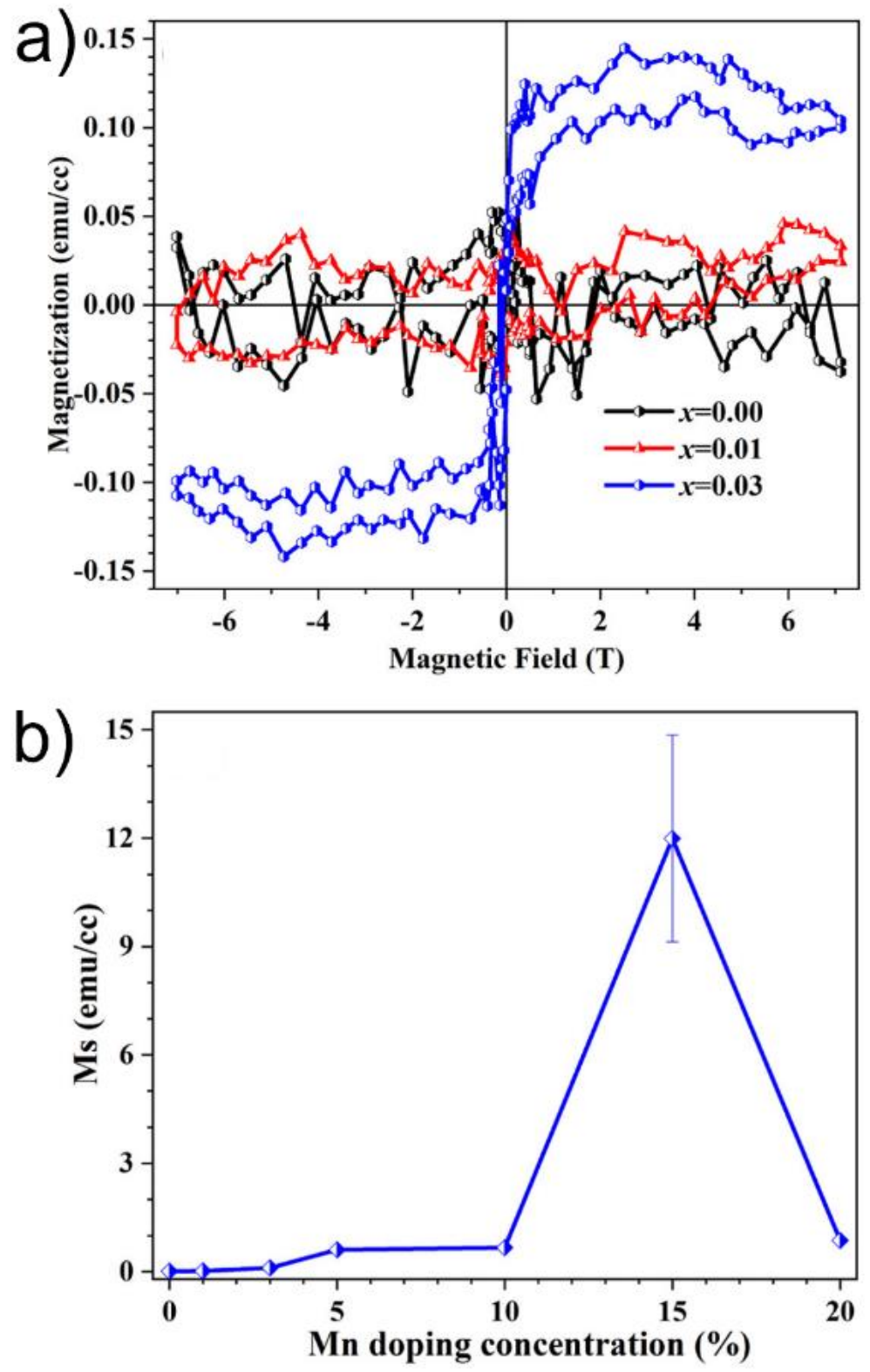

*Fig. 7. a) SQUID M−H curves for the $MAPb_{1-x}Mn_xI_3$ films b) Dependence of saturation magnetization ($M_s$) on Mn doping concentration in 450 nm thick $MAPb_{1-x}Mn_xI_3$ films. Reprinted with permission from ref. [116] Copyright © 2021 American Chemical Society.*

This section explores applications that rely directly on the magnetic properties of HOIPs, from magneto-optoelectronic devices to spintronics, demonstrating the transformative impact of transition metal doping on HOIP functionality.

### 5.1. Magnetic control of photocurrent

The transformation of diamagnetic $(CH_3NH_3)PbI_3$ into a ferromagnet through Mn doping in varying proportions has been explored by Ren et al.[133] By substituting some $Pb^{2+}$ atoms with $Mn^{2+}$ to create $(CH_3NH_3)Pb_{1-x}Mn_xI_3$, with molar ratios x=0, 0.05, 0.1, and 0.2, they made a doped thin film on a glass substrate using solution processing, resulting in the device shown in Fig. 8a. The magnetic properties of the material were then analyzed with a superconductive quantum interference device (SQUID).

To investigate the influence of magnetic fields on photoelectric performance, $(CH_3NH_3)Pb_{0.8}Mn_{0.2}I_3$ was chosen and compared to undoped $(CH_3NH_3)PbI_3$. The relative change in short-circuit photocurrent ($MI_P$) was defined as the difference between the photocurrent measured at a specific magnetic field ($I_P(H)$) and that at

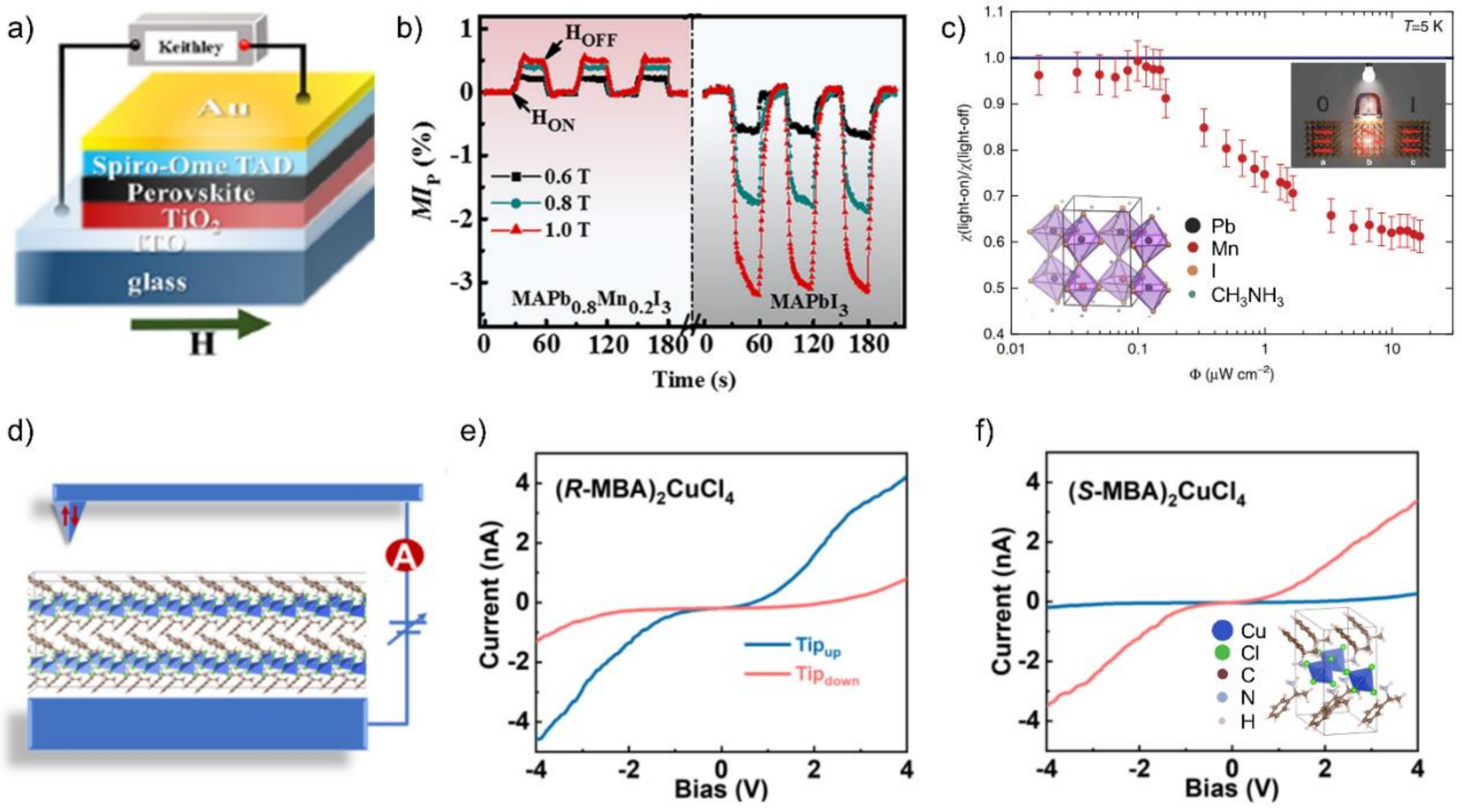

*Fig. 8. a) Schematic diagram of the photovoltaic device fabricated to study the magnetic dependence of the photocurrent on $MAPb_{1-x}Mn_xI_3$ (x= 0, 0.05, 0.1, 0.2). b) Relative change of short-circuit photocurrent in time under different external magnetic field values, and comparison of the behaviour between the ferromagnetic $MAPb_{1-x}Mn_xI_3$ and the non-magnetic $MAPbI_3$. Reprinted with permission from ref. [136]. Copyright © 2020, American Chemical Society c) Spin susceptibility over a broad range of illumination intensities. The ferromagnetic order decreases monotonically after a threshold, leading to the melting of the ferromagnetism. The upper inset represents the working principle of a RAM based on this phenomenon. The lower inset depicts the crystal structure of $MAPbI_3$. Ref. [63] Copyright © 2016, The Author(s) d) Schematic diagram of the*

*magnetic conductive-probe atomic force microscopy set-up used for the study of spin filtering along the c-axis of (R/S-MBA)$_2$CuCl$_4$. e-f) Current intensity as a function of applied voltage for each enantiomer, under the probe magnetized in the upward (blue) and downward (red) directions, showing the opposite response. The inset in figure f represents the crystal structure of (R/S-MBA)$_2$CuCl$_4$. Ref. [65] © 2021 Wiley-VCH GmbH.*

zero magnetic field ($I_P(0)$), divided by the photocurrent at zero magnetic field ($I(0)$), as illustrated in Fig. 8b. This relationship is expressed by the equation:

$$MI_P = \frac{I_P(H) - I_P(0)}{I(0)} \cdot 100\%$$

In experiments, an increase in $MI_P$ of 0.5% was observed in the Mn-doped perovskite, while the undoped $MAPbI_3$ exhibited a 3.3% decrease. This enhancement in photocurrent in the doped perovskites is attributed to a reduction in photoresistance, a result of Mn ions' spin alignment under the magnetic field, which intensifies the $d^5$-$e_g$ double exchange and facilitates electron hopping.

### 5.2. Optically switched magnetism for RAMs

In another study on Mn-doped $(CH_3NH_3)PbI_3$, Náfrádi et al.[63] increased the Mn content, substituting 10% of the $Pb^{2+}$ atoms by $Mn^{2+}$, and investigated the material's magnetic properties using electron spin resonance (ESR). Their results indicate that illumination at wavelengths shorter than the bandgap (830 nm) reduces ferromagnetic order, effectively "melting" the magnetic structure, as illustrated in Fig. 8c. This phenomenon may result from the interplay between superexchange and light-induced Ruderman-Kittel-Kasuya-Yosida (RKKY) interactions. While superexchange stabilizes magnetic order in the absence of light, illumination introduces conduction electrons that alter spin alignment through the RKKY mechanism.

The RKKY interaction, a phenomenon where magnetic moments in a material are indirectly coupled through conduction electrons, can create oscillating magnetic interactions that vary in strength and orientation depending on electron density.

This effect holds promise for magnetic memory applications, allowing magnetic bits to be written and rewritten via light illumination. By disrupting the existing magnetic order with light, a small, localized field can then be applied to establish a new magnetic moment direction (inset of Fig. 8c). This mechanism paves the way for rapid, low-power optical control of spin states.

### 5.3. Spin filtering

Spin filtering is a key process in creating spin-polarized currents, which is fundamental for spintronic devices such as spin valves and magnetic tunnel junctions. Lu et al.[65] study spin filtering on thin films of (R/S-MBA)$_2$CuX$_4$, X = Cl, Br, using magnetic conductive-probe atomic force microscopy. Their findings indicate that the right-handed (R) enantiomer selectively filters out spin-down charge carriers, while the left-handed (S) enantiomer filters out spin-up carriers, as shown in Fig. 8e-f.

This spin filtering effect is attributed to chiral-induced spin selectivity (CISS), which arises from the chiral structure of the HOIPs combined with a high degree of spin polarization. The results align with CISS theory, which suggests that the chiral molecular structure can influence electron spin orientation, thereby enabling selective spin filtering.

### 5.4. Magnetic modulation of photoluminescence emission

In a comprehensive study, Neumann et al.[64] examined the magneto-optical behavior of the paramagnetic 2D Mn-doped $(C_6H_5(CH_2)_2NH_3)_2PbI_4$ (Mn:$PEA_2PbI_4$) using temperature- and magnetic field-dependent photoluminescence spectroscopy. Their work focused on circularly polarized photoluminescence (CPL) driven by spin-dependent exciton recombination. Unlike pristine $PEA_2PbI_4$, Mn:$PEA_2PbI_4$ exhibits CPL that is responsive to applied magnetic fields. Specifically, the intensity of left- and right-handed CPL peaks shift in opposite directions when subjected to positive and negative magnetic fields, as shown in Fig. 9a-b. Additionally, the CPL degree increases with higher Mn concentrations, reaching up to 13% at 1% Mn doping (Fig. 9c).

Two mechanisms are proposed to explain this CPL behavior. First, variations in carrier excitation rates and dark exciton formation may differ across spin states, leading to a spin population imbalance during recombination that manifests as CPL. Alternatively, Mn dopants might introduce an additional radiative recombination pathway. In this pathway, a normally forbidden transition becomes accessible if Mn spin flips simultaneously with exciton recombination. This pathway shows increased efficiency for the exciton spin state aligned with Mn spins under an external magnetic field, resulting in CPL through a similar spin population imbalance.

In a complementary study, the modulation of photoluminescence (PL) in the 2D HOIP $(CH_3NH_3)_2MnCl_4$ under varying magnetic fields and temperatures has been systematically studied by Zhang et al.[66] using detailed PL spectroscopy measurements. They demonstrated that the PL characteristics of $(CH_3NH_3)_2MnCl_4$ are highly dependent on the magnetic ordering of $Mn^{2+}$ ions, which undergo transitions between antiferromagnetic canted ferromagnetic, and SF states as temperature and external magnetic fields vary.

Below 47 K, the $Mn^{2+}$ spins exhibit antiferromagnetic ordering that suppresses PL by facilitating energy transfer to non-radiative states through superexchange interactions (Fig. 9d). In contrast, as an external magnetic field is applied, antiferromagnetic ordering is progressively disrupted, and a SF transition occurs at approximately 3.5 T, leading to partial ferromagnetic alignment and a pronounced increase in PL intensity (Fig. 9e). This enhancement is attributed to the suppression of non-radiative energy transfer, allowing more efficient radiative recombination. Moreover, at very high magnetic fields, the complete alignment of $Mn^{2+}$ spins into a ferromagnetic state further increases PL intensity and induces a redshift in the PL peak energy.

Two key mechanisms are identified to explain the observed magneto-optical effects. First, antiferromagnetic ordering favors non-radiative energy loss pathways, reducing PL intensity, while ferromagnetic ordering mitigates these losses, enhancing optical transitions. Second, the external magnetic field alters the selection rules governing *d-d* transitions in $Mn^{2+}$ ions, facilitating normally forbidden transitions by breaking spin degeneracy. As the magnetic field increases, energy levels split further, leading to shifts in the emission spectrum and different trends in the PL intensity, which vary depending on the temperature. For example, in the range 60-80 K the intensity of PL reduces as the magnetic field increases, while for temperatures lower than 30 K, the PL intensity is enhanced. It is the competitive interplay between antiferromagnetic and ferromagnetic orderings, together with electron-phonon interactions, what dictates the temperature and field-dependent PL properties of $(CH_3NH_3)_2MnCl_4$.

This work highlights the potential of magnetic-field-modulated PL in $Mn^{2+}$-based perovskites for applications in magneto-optical and spin-photonic devices, where fine-tuning of emission properties is critical.

**5.5. Prospects for potential applications**

The exploration of HOIPs in multifunctional devices is gaining attraction, with recent studies highlighting the potential of magneto-optoelectronic, magnetocaloric, and multifunctional applications due to their unique properties.

Regarding magneto-optoelectronics, Zhang et al.[138] synthesized $(C_4H_{10}N)MnBr_3$ (pyrrolidinium as organic cation) and studied its electrical, magnetic and optical properties. They found that this HOIP is ferroelectric, shows weak FM and exhibits intense red PL under UV excitation. The combination of multiferroicity and PL emission opens the door to fabricating multifunctional magneto-optoelectronic devices.

In the field of magnetocaloric materials for cooling devices, Septiany et al.[139] investigated the magnetocaloric properties of the ferromagnetic $(PMA)_2CuCl_4$. They determined the performance of the material via the relative cooling power (RCP), which can be obtained by determining the magnetic entropy change as a function of temperature. They found that under a magnetic field of 7 T the RCP reaches 47.8 J/kg, a value large enough to consider its application of cooling devices. A similar study was performed by Ma et al.[140] on $(CH_3NH_3)_2CuCl_4$ showing a large magnetocaloric effect as well. Other studies focus on the HOIPs $(CH_3(CH_2)_{11}NH_3)_2Cu(Br_{1-x}Cl_x)_4$, $(C_6H_9(CH_2)_2NH_3)_2Cu(Br_{1-x}Clx)_4$[141] and $(CH_3CH_2NH_3)_2CoCl_4$[142].

As for multifunctional devices, several research groups have demonstrated that 2D HOIPs can exhibit diverse properties, making them suitable for a variety of applications. For example, Sun et al.[143] synthesized and studied the 2D HOIP $(PED)CuCl_4$ and $(BED)_2CuCl_6$ (being PED – N-phenylethylenediammonium and BED – N-benzylethylenediamine), which show strong FM below $T_C$ = 13K. They also found reversible thermochromism, which shows applications in thermal sensors; and a six-order-of-magnitude conductivity change in $(BED)_2CuCl_6$ upon temperature change, opening the door to designing multifunctional devices.

Likewise, Xiong et al.[144], using the chiral molecule $C_6H_{15}ClNO$ (CTA=3-chloro-2-hydroxypropyltrimethylammonium), synthesized the chiral HOIPs $(S\text{-}CTA)_2CuCl_4$ and $(R\text{-}CTA)_2CuCl_4$. Their extensive investigation reveals properties such as ferroelasticity, thermochromism, and chirality-induced effects, along with interactions among these characteristics. Notably, they identify seven physical channel switches, indicating the potential to induce transitions or changes in the properties of these perovskites through external stimuli. While their study does not explicitly address the magnetic properties of these materials, the presence of Cu suggests that they could exhibit magnetic behavior, thereby opening avenues for further exploration in applications like spintronic devices, memory storage, sensors, and smart windows, all of which could benefit from magnetism. Another example are the 2D polar ferromagnets $(3\text{-}ClbaH)_2CuCl_4$, $(4\text{-}ClbaH)_2CuCl_4$ and $(2\text{-}ClbaH)_2CuCl_4$ ($ClbaH^+$= chlorobenzylammonium) reported by Han et al.[142] These HOIPs not only present FM, but also second harmonic generation and polarity, so they are candidates for multifunctional devices. Other perovskites that are proved, experimentally and/or theoretically, to have multiferroicity are $(C_6H_5CH_2NH_3)CuCl_4$[143], $(CH_2NH_3)_2[FeCl_4]$, $(CH_3NH_3)_2[FeCl_4]$[92] and $(CH_3CH_2CH_2NH_3)_2FeCl_4$.[58]

In summary, magnetic HOIPs present exciting opportunities for advancing applications in spintronics, magneto-optoelectronics, and multifunctional devices due to their unique magnetic properties and tunability. Transition metal doping, in particular, has enabled promising proof-of-concept applications that leverage magnetism within perovskite frameworks, demonstrating the potential for high-impact innovations. Notably, as discussed in Section 4.1, pure transition-metal-based HOIPs exhibit stronger magnetic interactions compared to doped perovskites, owing to superexchange mechanisms. This distinction could be especially relevant for future applications that require more robust magnetic coupling. However, despite their potential, most application-focused research has centred on doped HOIPs. Additionally, research on applications is in early stages and many challenges need to be addressed to transition these materials from the laboratory to real-world devices. Continued investigation into their magnetic behaviours and interaction mechanisms will be critical for realizing the full potential of these materials in future technological applications.

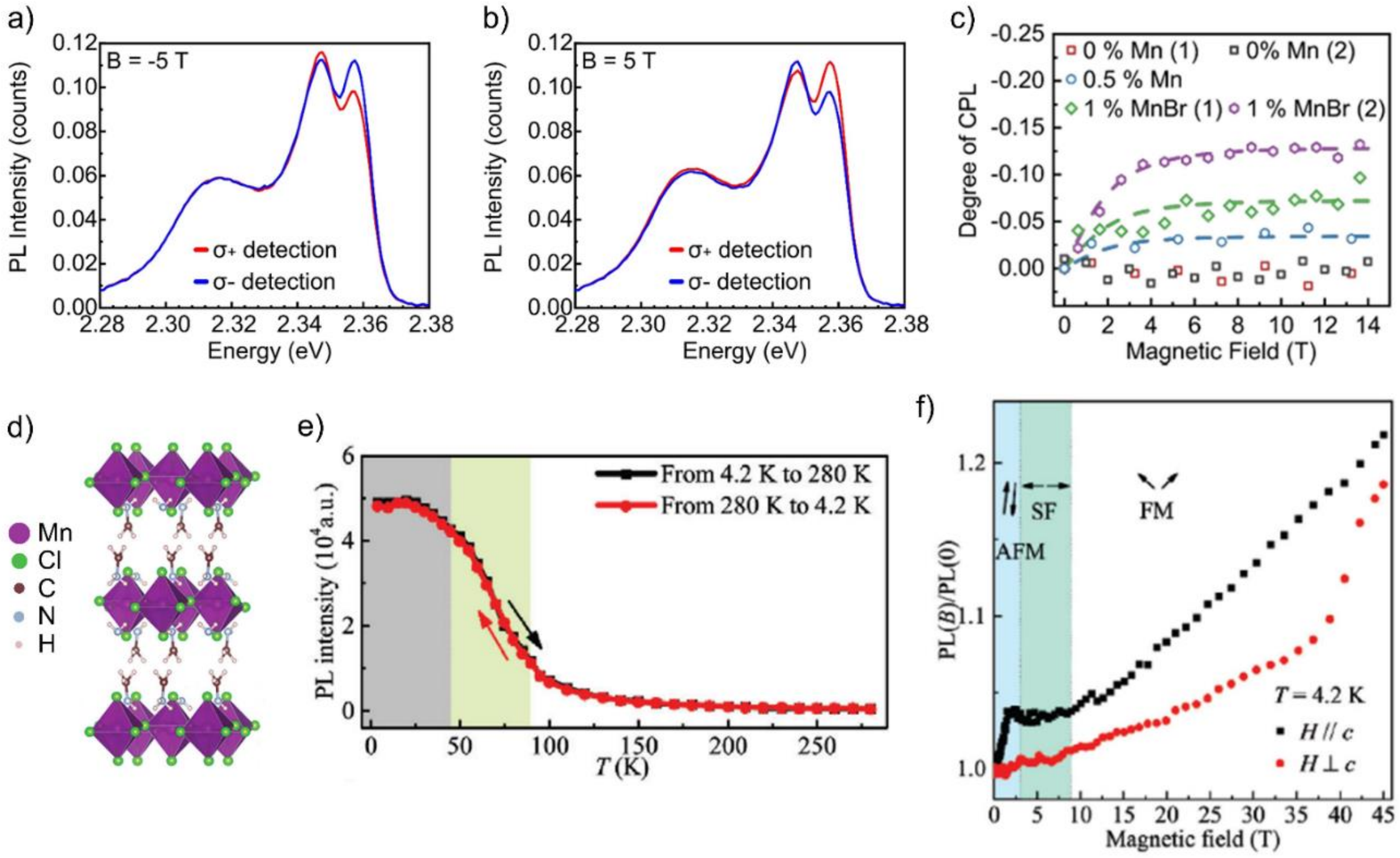

*Fig. 9. a-b) Low-temperature (4 K) PL spectra of Mn:$PEA_2PbI_4$ under a negative and a positive magnetic field respectively, showing opposite response between the spectra collected for right-handed and left-handed detection for each case. c) Behavior of the degree of CPL with magnetic field as a function of the dopant content. The degree of CPL increases with increasing Mn content, while shows no variation for undoped $PEA_2PbI_4$. Ref. [64] Copyright © 2021, The Author(s) d) Representation of the crystal structure of $(CH_3NH_3)_2MnCl4$. e) Temperature dependence of the PL intensity on heating and cooling, showing the different behavior of PL below and above 47 K. f) Change of PL with magnetic field at 4.2 K for in-plane and out-of-plane configurations, highlighting the AFM to SF and canted FM transitions. Ref. [66] © 2024 Wiley-VCH GmbH.*

## 6. Conclusions and outlook

Magnetic HOIPs stand out as a promising class of materials for next-generation magnetic and multifunctional devices. This review has highlighted the unique magnetic properties and structural flexibility of HOIPs and demonstrated the significant potential to tune these materials through chemical and structural manipulation. Despite the progress made, challenges remain in achieving long-term stability, scalable synthesis, and on-demand control over magnetic properties, which are essential for advancing HOIPs into practical device applications. Looking forward, there are clear opportunities and challenges to address in the next future:

- Stability enhancement: The ambient instability of HOIPs poses a significant barrier to their use in devices, and several solutions have been proposed namely for the most studied Pb ones such as surface modifications, encapsulation, or lattice substitutions. But these approaches should be explored in magnetic HOIPs to make them viable for long-term applications. This is particularly important for devices that require consistent performance in diverse environments.
- Fine-tuning magnetic properties: While the magnetism of Cu-based HOIPs has been deeply investigated, this review highlights that Mn, Cr, and Fe-based HOIPs also show promising, tunable characteristics, which yet remain underexplored. Mn-based HOIPs, for example, offer multifunctional potential since Mn in octahedral coordination can produce red photoluminescence, which can be chirality-dependent, particularly with the inclusion of chiral organic molecules. Likewise, Fe-based HOIPs present further possibilities, such as incorporating ferroelectric properties. Additionally, double perovskites increase the opportunities for tuning the properties of the HOIP by e.g. varying each metal cation differently. Further investigation into the effects of dimensionality, organic cation selection, halide composition, as well as expanding research on these underexplored systems, could enable on-demand magnetic characteristics for specific device requirements, facilitating more versatile applications in spintronics, magneto-optics, and optoelectronics. In this direction, the increase of the Curie temperature and the study of new phenomena already highlighted in metal halide 2D magnets such as spin dynamics and topological features are two key points to explore in the future.
- Advanced synthesis techniques: Here we focus on synthesis protocols for obtaining single bulk crystals, but integrating these materials in current technologies and devices demands large-scale, reproducible, and efficient methods that are able to produce good quality layers over defined areas. Improvements in solution and vapor-phase approaches are key to achieving commercially feasible, high-quality and crystal-oriented films at a lower cost, making HOIPs a feasible option for device integration.
- From an application perspective, magnetic HOIPs offer immense potential across multiple domains, such as magnetic data storage, magneto-optics, sensors, spin filters, and spatial light modulators.[10,146] However, many current studies focus on the fundamental properties of HOIPs, with relatively few reporting fully realized devices. To bridge this gap, more research on lab-scale prototypes is needed, as well as deeper insights into the underlying physics, such as spin-dependent photo-physics and exciton polarization control.[64] By addressing these scientific and engineering challenges, HOIPs could become key materials for low-power, magnetically controlled devices, marking a significant advancement in spin-based and multifunctional technologies.

## Author contributions

Yaiza Asensio, Lucía Olano-Vegas: Conceptualization, Investigation, Visualization, Writing - original draft, review & editing. Samuele Mattioni: Conceptualization,

Investigation, Writing - original draft. Marco Gobbi, Fèlix Casanova: Funding acquisition, Project administration, Resources, Supervision, Writing - review & editing. Luis E. Hueso: Conceptualization, Funding acquisition, Project administration, Resources, Supervision, Writing - review & editing. Beatriz Martín-García: Conceptualization, Funding acquisition, Project administration, Resources, Supervision, Investigation, Visualization, Writing - original draft, review & editing.

**Conflicts of interest**

There are no conflicts to declare.

**Data availability**

No primary research results, software or code have been included and no new data were generated or analysed as part of this review.

**Acknowledgements**

This work is supported under Projects PID2021-122511OB-I00 and PID2021-128004NB-C21 and under the María de Maeztu Units of Excellence Programme (Grant CEX2020-001038-M) funded by Spanish MICIU/AEI/10.13039/501100011033 and by ERDF/EU. Additionally, this work was carried out with support from the Basque Science Foundation for Science (IKERBASQUE), concretely, B.M.G. thanks IKERBASQUE HYMNOS project. This work is also supported by the FLAG-ERA grant MULTISPIN, by the Spanish MICIU/AEI/10.13039/501100011033 and European Union NextGenerationEU/PRTR with grant number PCI2021-122038-2A. Y.A. and L.O-V. thank the funding from Spanish MICIU/AEI/10.13039/501100011033 and ESF+ (PhD grants PRE2021-099999 and PRE2022-104385, respectively). B.M.-G. and M.G. thanks support from "Ramón y Cajal" Programme by the Spanish MICIU/AEI/10.13039/501100011033 and European Union NextGenerationEU/PRTR (grant nos. RYC2021-034836-I and RYC2021-031705-I, respectively). Authors thank Dr. E. Goiri Little (Nanodevices group – CIC nanoGUNE) for reading and revising the manuscript.